\documentclass[12pt]{article}
\pdfoutput=1
\usepackage[centertags]{amsmath}
\allowdisplaybreaks[1]
\usepackage{array,multirow}
\numberwithin{equation}{section}
\usepackage{amssymb}
\usepackage{float}
\usepackage{graphicx}
\usepackage{color}
\usepackage{relsize}

\usepackage{epsfig}

\def\m{\mu}

\def\Or[#1]{{\text{O}}\left({#1}\right)}
\def\dotl[#1,#2]{\left\langle #1, #2 \right\rangle}
\def\dotlb[#1,#2]{[ #1, #2 ]}
\def\dotp[#1,#2]{(#1) \cdot (#2)}
\def\aff[#1,#2]{\hat{#1}(#2)}
\def\n4sym{{\cal N}=4 SYM}
\def\>{\rangle}
\def\<{\langle}
\def\weight[#1,#2,#3]{\{(#1),#2,#3\}}
\def\ads[#1]{$\text{AdS}_{#1}$}

\newcommand{\ba}{\begin{eqnarray}}
\newcommand{\ea}{\end{eqnarray}}
\newcommand{\be}{\begin{eqnarray}}
\newcommand{\ee}{\end{eqnarray}}
\newcommand{\bq}{\begin{equation}}
\newcommand{\eq}{\end{equation}}
\newcommand{\benn}{\begin{equation*}}
\newcommand{\eenn}{\end{equation*}}
\newcommand{\bi}{\begin{itemize}}  
\newcommand{\ei}{\end{itemize}}

\newcommand{\CO}{{\cal O}}

\newcommand{\CV}{{\cal V}}
\newcommand{\nn}{\nonumber}

\newcommand\oo\infty
\newcommand\s\sigma
\newcommand\de\delta
\newcommand\De\Delta

\newcommand\f\phi
\newcommand\g\gamma
\newcommand\x\times

\usepackage{jheppub}

\makeatletter
\def\@fpheader{\vspace{-.1cm}}
\makeatother

\title{A Numerical Approach to Virasoro Blocks \\ and the Information Paradox}

\author{Hongbin Chen,}
\author{Charles Hussong,}
\author{Jared Kaplan, and}
\author{Daliang Li}
\affiliation{Department of Physics and Astronomy,  Johns Hopkins University, \\
Charles Street, Baltimore, MD 21218, U.S.A.}

\abstract{ 
We chart the breakdown of semiclassical gravity by analyzing the Virasoro conformal blocks to high numerical precision, focusing on the heavy-light limit corresponding to a light probe propagating in a BTZ black hole background.  In the Lorentzian regime,  we find empirically that the initial exponential time-dependence of the blocks transitions to a universal $t^{-\frac{3}{2}}$ power-law decay.  For the vacuum block the transition occurs at $t \approx \frac{\pi c}{6 h_L}$, confirming analytic predictions.   In the Euclidean regime, due to Stokes phenomena the naive semiclassical approximation fails completely in a finite region enclosing the `forbidden singularities'.  We emphasize that limitations on the reconstruction of a local bulk should ultimately stem from distinctions between semiclassical and exact correlators. }

\arxivnumber{}

\begin{document}

\maketitle
\flushbottom
 
\section{Introduction and Summary}

Many of the most challenging conceptual problems in theoretical physics were only resolved after physicists discovered how to `shut up and calculate' a large variety of observables to high precision.  For example, our modern understanding of quantum field theory was only developed after the physics community had decades of experience with perturbative calculations.  And it is hard to imagine how decoherence could have been understood without the temporary crutch provided by the Copenhagen interpretation and its instrumental approach to the Born rule.

Though we have struggled with the black hole information paradox for decades,  major progress has been possible through the development of AdS/CFT.  Resolving the information paradox in AdS/CFT will require a precise understanding of bulk reconstruction and its limitations.  Although reconstruction presents thorny conceptual problems, the limitations on reconstruction  should ultimately stem from discrepancies between the predictions of gravitational effective field theory in AdS and conformal field theory. This means that to make progress, it will be crucial to be able to directly compare the approximate correlation functions of bulk EFT and the exact correlators of the CFT.  

AdS$_3$/CFT$_2$ may provide the best opportunity for such comparisons.
Many features of quantum gravity in AdS$_3$ can be understood  `kinematically' as a consequence of the structure of the Virasoro algebra.  To be specific, the Virasoro conformal blocks have a semiclassical large central charge limit that precisely accords with expectations from AdS$_3$ gravity, reproducing the physics of light objects probing BTZ black holes.  In the semiclassical approximation, the Virasoro blocks exhibit information loss in the form of `forbidden singularities' and exponential decay at late times \cite{Fitzpatrick:2014vua, Fitzpatrick:2015zha, Fitzpatrick:2015foa, Fitzpatrick:2015dlt, Fitzpatrick:2016mjq, Anous:2016kss}.  Moreover, these problems can be partially addressed by performing explicit analytic calculations \cite{Fitzpatrick:2016ive}.  The  blocks can also be computed directly from AdS$_3$ \cite{KrausBlocks, Hijano:2015qja, Besken:2016ooo, Alkalaev:2015wia, Alkalaev:2015lca, Hulik:2016ifr, Fitzpatrick:2016mtp, Besken:2017fsj}.

In this work we will  investigate the discrepancies between semiclassical gravity and the exact CFT by computing the Virasoro blocks numerically to very high precision.  This is possible via a slightly non-trivial implementation of the Zamolodchikov recursion relations \cite{ZamolodchikovRecursion, Zamolodchikovq, Zamolodchikov:1986gh}.  We discuss the blocks and the algorithm in detail in section \ref{sec:BlocksZRR} and appendix \ref{app:ZRecursion}.  For the remainder of the introduction we will explain the physics questions to be addressed and summarize the results.

\subsection*{When is the Semiclassical Approximation Valid?}

The Virasoro conformal blocks have a semiclassical limit.  CFT$_2$ correlators can be written in a Virasoro block decomposition as
\be 
\< \CO_1(0)  \CO_2(z) \CO_3(1) \CO_4(\infty) \> = \sum_{h, \bar h} P_{h, \bar h} \CV_{h_i, h,c}(z) \CV_{\bar h_i, \bar h,c}(\bar z)
\ee
the holomorphic Virasoro blocks $\CV_{h_i,h,c}$ depend on the holomorphic dimensions $h_i$ of the primary operators $\CO_i$, on an intermediate primary operator dimension $h$, and on the central charge $c$.  A semiclassical limit emerges when $c \to \infty$ with all $h_i/ c$ and $h/c$ fixed; the blocks take the form 
\be \label{eq:SemiclassicalForm}
\CV = e^{-\frac{c}{6} f \left( \frac{h_i}{c},\frac{h}{c}, z \right)}  
\ee
It is natural to ask about the range of validity of this approximation -- how does it depend on the kinematic variable $z$ and the ratios $h_i/c$ and $h/c$?

One reason to ask is simultaneously speculative and pragmatic -- one might like to know if it is possible to explore AdS$_3$ quantum gravity in the lab by engineering an appropriate CFT$_2$  (for a concrete idea see \cite{Plamadeala:2014roa}).  But gravity will only be a good description if the semiclassical limit provides a reasonable approximation at accessible values of $c$.  Unfortunately, even in the semiclassical limit the Virasoro blocks are not known in closed form for general parameters.  But we can partially test the validity of this limit by computing $\frac{c_2}{c_1} \frac{\log \CV(c_1)}{\log \CV(c_2)}$ for $c_2 \approx c_1$, as this ratio will be $1$ when the semiclassical limit holds.  We plot this ratio in figure \ref{fig:RatioOfLogV}, which shows that the blocks adhere to the semiclassical form of equation (\ref{eq:SemiclassicalForm}) remarkably well (up to an important caveat to be discussed later).

We would also like to understand if the semiclassical limit breaks down in specific kinematic regimes associated with quantum gravitational effects in AdS$_3$.  This is what we will explore next.

\subsubsection*{Information Loss and OPE Convergence in a New Regime}

\begin{figure}
\centering
\includegraphics[width=0.75\textwidth]{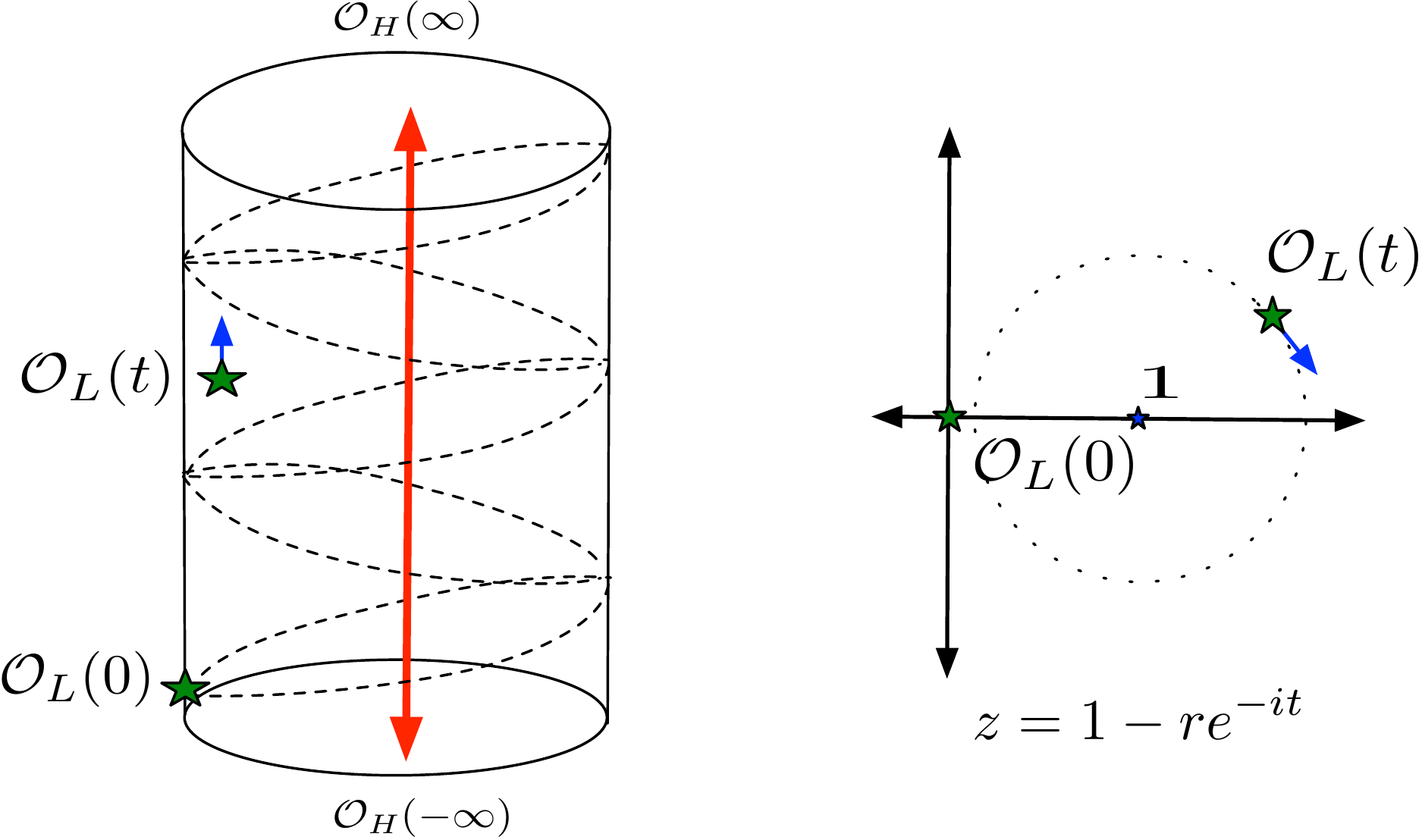}	
\caption{This figure suggests the analytic continuations necessary to obtain a heavy-light correlator with increasing (Lorentzian) time separation between the light operators.  We take $r \lesssim 1$ to avoid singularities on the lightcones displayed on the left; one can also use $r$ as a proxy for a Euclidean time separation between the light operators. }
\label{fig:TimelikeSeparation}
\end{figure}

Information loss can be probed using the correlators of light operators in a black hole background \cite{Maldacena:2001kr}, as illustrated in \ref{fig:TimelikeSeparation}.  When computed in a BTZ or AdS-Schwarzschild geometry, these correlators decay exponentially as we increase the time separation $t$ between the light operators, even as we take $t \to \infty$.  This behavior represents a violation of unitarity for a theory with a finite number of local degrees of freedom on a compact space.   Thus it is interesting to see how it is resolved by the exact CFT description.

As a first step, one would like to understand how unitary CFTs are able to mimic bulk gravity, including the appearance of information loss.  This arises from the heavy-light large central charge approximation \cite{Fitzpatrick:2016ive, Anous:2016kss, Fitzpatrick:2016mjq} of the Virasoro blocks.  Thus it seems to be very universal, as it is largely independent of CFT data.   The next step is to understand how finite $c$ physics corrects this approximation, and what CFT data are involved in resolving information loss.

It is useful to think about the Fourier representation of both the correlator and the individual Virasoro blocks.  For the full correlator this is
\be
\< \CO_H(\infty) \CO_L(t) \CO_L(0) \CO_H(-\infty) \> = \int d E \,  \lambda^2(E) e^{i E t}
\ee
where $\lambda(E)$ is the OPE coefficient density and we have taken $z = 1 - e^{-it}$ to study Lorentzian time separations between the light probe operators.  At large $t$ we probe the fine structure of $\lambda(E)$, which means that the least analytic features of $\lambda(E)$ dominate the late time limit.  Practically speaking, this means that the very late time limit probes the discrete nature of the spectrum, and we become sensitive to the fact that $\lambda^2(E)$ is a sum of delta functions.  At early or intermediate times we only discern the coarse features of $\lambda(E)$.  

There are at least five different timescales associated with black holes in AdS/CFT.  The inverse temperature $\beta = \frac{2 \pi}{|\alpha_H|}$ where $\alpha_H \equiv \sqrt{1- \frac{24 h_H}{c}}$  sets the shortest relevant scale, where $h_H$ is the holomorphic dimension of $\CO_H$.  The scale $\beta \log c$ estimates the time it takes for infalling matter to be scrambled \cite{Hayden:2007cs, Maldacena:2015waa}.  At times of order the entropy $S = \frac{\pi^2 c}{3 \beta}$, heavy-light correlators cease their exponential decay; this is also the evaporation timescale for black holes in flat spacetime.  We expect that the typical energy splittings among neighboring eigenstates to be of order $e^{-S}$, which means that at times of order $e^S$ we will be sensitive to the discreteness of the spectrum.  Finally, on timescales of order $e^{e^S}$ the phases of the eigenestates can come back into approximate alignment, leading to recurrences.


As discussed in detail in section \ref{sec:LateTime}, what we find is that the Virasoro blocks with $h_L < \frac{c}{24} < h_H$ behave very differently at early and late times, as was presaged by analytic results \cite{Fitzpatrick:2016ive}:
\begin{itemize}
\item The blocks with intermediate operator dimension $h \lesssim \frac{c}{24}$ are well-described by their semiclassical limit \cite{Fitzpatrick:2014vua, Fitzpatrick:2015zha, Fitzpatrick:2016mjq} for
\be
t \lesssim t_D \equiv \frac{\pi c}{6 h_L}
\ee 
When $h > \frac{c}{24}$ the blocks are also well-described by the semiclassical limit at early times, but we do not have a precise formula quantifying `early'.
\item  Heavy-light blocks with $h \gtrsim h_H$ initially grow, as was found from a semiclassical analysis \cite{Fitzpatrick:2016mjq}.  We find that they reach a maximum 
\be \label{eq:IntroVmaxtmax}
|\CV|_{\max} \approx 16^{h - \frac{c-1}{24}}  \left( \frac{h}{c} \right)^{-\frac{5}{2} h_H} \ \ \ \ \mathrm{at} \ \ \ \ t_{\max} \approx A_{t}\, \sqrt{\frac{24 h}{c} - 1}
\ee
and then subsequently decay. The factor $\frac{5}{2}$ comes from empirical fits; the function $A_{t}(\frac{c}{h_H})$ is always order one and is approximately linear in $\frac{c}{h_H}$.  Other sub-leading behavior is discussed in section \ref{sec:LateTime}.
\item Numerical evidence indicates that all heavy-light Virasoro blocks decay as
\be
\left|\CV(t \gg t_D)\right| \propto t^{-\frac{3}{2}}
\ee
at late times, independent of $h$, as long as $h_H > \frac{c}{24} > h_L$.  We present evidence that this decay persists beyond the exponentially long timescale $\sim e^S$, so we believe that it represents the true asymptotic behavior of the heavy-light blocks.
\end{itemize}
From the point of view of the  $\frac{1}{c} \propto G_N$ expansion, the universal late-time power-law decay comes from non-perturbative effects.  If this behavior persists to all times, as our empirical evidence indicates, then the late time behavior of CFT$_2$ correlators must come from an infinite sum over Virasoro blocks in the heavy-light channel.\footnote{In the $\CO_H \CO_L \to \CO_H \CO_L$ OPE channel, the late time behavior can be understood from the discreteness of the spectrum, without including states with energies $E \gg h_H$.  It appears that in the $\CO_L \CO_L \to \CO_H \CO_H$ channel one needs to include states of arbitrarily high energy.}

From a pure CFT perspective, the late Lorentzian time behavior  represents a new limit in which the bootstrap may be analytically tractable.  Most analytic bootstrap results, including the Cardy formula \cite{Cardy:1986ie}, OPE convergence \cite{Pappadopulo:2012jk}, and the lightcone OPE limit \cite{Fitzpatrick:2012yx, KomargodskiZhiboedov} arise in a similar way.  In fact, because the expansion of CFT$_2$ correlators in the uniformizing $q$-variable converges everywhere, including in deeply Lorentzian regimes, it affords the opportunity to explore many new `analytic bootstrap' limits.

\subsubsection*{Forbidden Singularities and Bulk Reconstruction}

\begin{figure}
\centering
\includegraphics[width=0.55\textwidth]{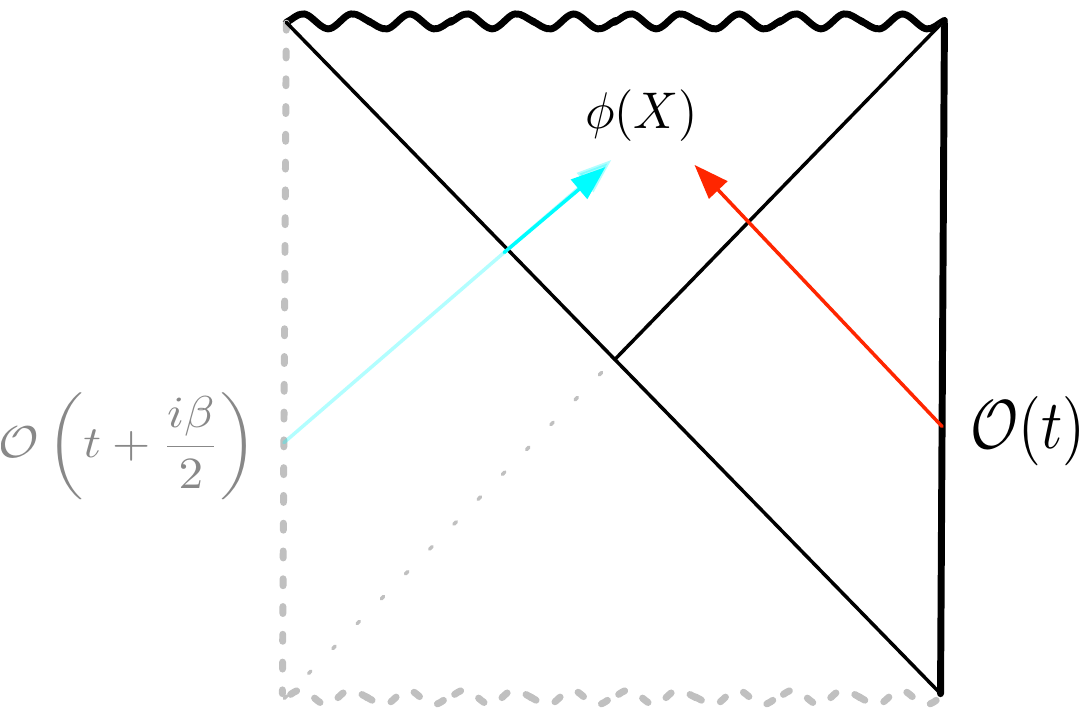}	
\caption{This figure shows the Penrose diagram for an energy eigenstate black hole in AdS, suggesting the role of ingoing and outgoing modes behind the horizon and their relationship with local CFT operators.  Analytic continuation provides a painfully naive but instrumentally effective method for studying correlators behind the horizon.  }
\label{fig:PenroseIngoingOutgoing}
\end{figure}

It is interesting to understand when exact CFT correlators differ markedly from predictions obtained from a semiclassical AdS description.   The late time regime we discussed above provides one example of this phenomenon.  As we discuss here and in section \ref{sec:ForbiddenSingularitiesandReconstruction}, there are also Euclidean regimes where the semiclassical approximation to the Virasoro blocks fails completely.

Correlation functions  in CFT$_2$ must be non-singular away from the OPE limits where local operators collide \cite{Maldacena:2015iua, Fitzpatrick:2016ive}.  The Virasoro conformal blocks must have this same property \cite{Fitzpatrick:2016mjq}.  But in the semiclassical approximation, the blocks develop additional `forbidden singularities' \cite{Fitzpatrick:2016ive} that represent a violation of unitarity.  These singularities are a signature of semiclassical black hole physics in AdS$_3$.  They arise because thermal correlators exhibit a Euclidean-time periodicity under $t \to t + i \beta$,  and so the OPE singularities have an infinite sequence of periodic images.  The exact Virasoro blocks  are not periodic, but in the semiclassical approximation they develop a periodicity at the inverse Hawking temperature $\beta = \frac{1}{T_H}$ associated with a BTZ black hole  in AdS$_3$.

By studying the Virasoro vacuum block in the vicinity of potential forbidden singularities, one can show that at finite $c$ the singularities are resolved in a universal way \cite{Fitzpatrick:2016ive} via an analytic computation.   This method predicts the kinematic regimes where non-perturbative effects should become important; it can be extracted from equation \ref{eq:2ndOrderVacDiffEq} and the results are displayed in figure \ref{fig:ContoursSigmaFunction}.  Thus it is interesting to investigate the divergence between the exact blocks and their semiclassical approximation more generally.  We study this question numerically in section \ref{sec:ForbiddenSingularitiesandReconstruction}. 

Discrepancies between exact and semiclassical CFT correlators near the forbidden singularities could have implications for the reconstruction of AdS dynamics.  Bulk reconstruction in black hole backgrounds is rather subtle \cite{Hamilton:2005ju, Bousso:2012mh, Leichenauer:2013kaa, Morrison:2014jha}, and perhaps requires some understanding of the analytic continuation of CFT correlators.  But  there is also a very simple and physical reason to expect that the analyic properties of correlators could have something to do with black hole interiors \cite{Hamilton:2005ju}.  

As emphasized by Raju and Papadodimas \cite{Papadodimas:2012aq, Papadodimas:2013jku}, a field operator behind the horizon consists of both ingoing and outgoing modes, but only the ingoing modes can be immediately associated with local CFT operators.  This issue is portrayed in figure \ref{fig:PenroseIngoingOutgoing}.  The analytic continuation of local operators by $t \to t + \frac{i \beta}{2}$ provides a naive, instrumental source for the outgoing modes.\footnote{This idea has significant problems.  Although it may be applied to single-sided black holes, which are our object of study, it cannot then apply to the case of eternal black holes involving two different entangled CFTs.  But even in the single-sided case, there is a problem because the ingoing and outgoing modes must commute, yet this property may fail when we use $\CO(t)$ and $\CO(t + \frac{i \beta}{2})$ for the ingoing and outgoing modes \cite{Papadodimas:2012aq, Almheiri:2013hfa}.  It can be imposed by fiat if we take an appropriate linear combination of correlators with different analytic continuations. But this seems to require a form of state-dependence.  We have discussed this procedure rather than e.g. mirror operators \cite{Papadodimas:2012aq, Papadodimas:2013jku} because it is easier to define in an unambiguous way.  We thank Suvrat Raju and Daniel Harlow for correspondence on these issues.}  Thus it is natural to ask whether the exact and semiclassical correlators differ significantly at $t + \frac{i \beta}{2}$, which is `halfway' to the first forbidden singularity.  

We will observe in section \ref{sec:ForbiddenSingularitiesandReconstruction} that the exact and semiclassical correlators behave very similarly at these points, though they seem to differ markedly both very near (within $\frac{1}{\sqrt{c}}$) and beyond the first forbidden singularity.  The results can be seen in figure \ref{fig:forbidden}.  As previously discussed \cite{Fitzpatrick:2016ive}, we expect that Stokes and anti-Stokes lines emanate from the locations of the forbidden singlarities, so that different semiclassical saddle points dominate in different regions of the $q$-unit disk.  
It appears that different saddles dominate as we cross the locations of the forbidden singularities, so that the naive semiclassical blocks (the saddles that dominate near $q = 0$) are not a good approximation beyond the first singularity.  In fact the semiclassical approximation appears to break down in a finite kinematic region, as shown in figure \ref{fig:ContoursSigmaFunction}.  Furthermore, the existence of such regions seems to depend in an essential way on the presence of a black hole, ie a state with energy above the Planck scale ($h_H > \frac{c}{24}$), as semiclassical/exact agreement is excellent when $h_H < \frac{c}{24}$, as we see in figure \ref{fig:NoBHContours}.

Perhaps future investigations will uncover bulk observables that are sensitive to Stokes phenomena in the large $c$ expansion of the Virasoro blocks.  We hope that the black hole information paradox can be understood with more precision and detail through such calculations.  This work takes steps in that direction by identifying new kinematic regimes where the semiclassical limit breaks down badly and by providing results for the correct non-perturbative Virasoro blocks.

\section{Kinematics, Convergence, and the Semiclassical Limit}
\label{sec:BlocksZRR}

A great deal of information about the behavior of CFT$_2$ correlation functions is encoded in the structure of the Virasoro conformal blocks.  We are interested in 4-pt correlators of primary operators, which can be written as
\be \label{eq:GenericVirasoroDefinition}
\< \CO_1(0)  \CO_2(z) \CO_3(1) \CO_4(\infty) \> = \sum_{h, \bar h} P_{h, \bar h} \CV_{h_i, h,c}(z) \CV_{\bar h_i, \bar h,c}(\bar z)
\ee
where the $P_{h, \bar h}$ are products of OPE coefficients.  The $\CV_{h_i, h,c}(z)$ are the holomorphic Virasoro blocks, which will be the main object of study in this work.    The blocks are uniquely fixed by Virasoro symmetry and depend only on the external dimensions $h_i$, the exchanged primary operator dimension $h$, and the central charge $c$.  Often it will convenient to write $z = \frac{4 \rho}{(1+\rho)^2}$ so that the full $z$-plane lies inside the $\rho$ unit circle \cite{Pappadopulo:2012jk}.  The Virasoro blocks are not known in closed form, but they can be computed order-by-order in a series expansion using  recursion relations.  We provide a brief summary here, leaving the details to appendix \ref{app:ZRecursion}. 

\begin{figure}
\centering
\includegraphics[width=0.85\textwidth]{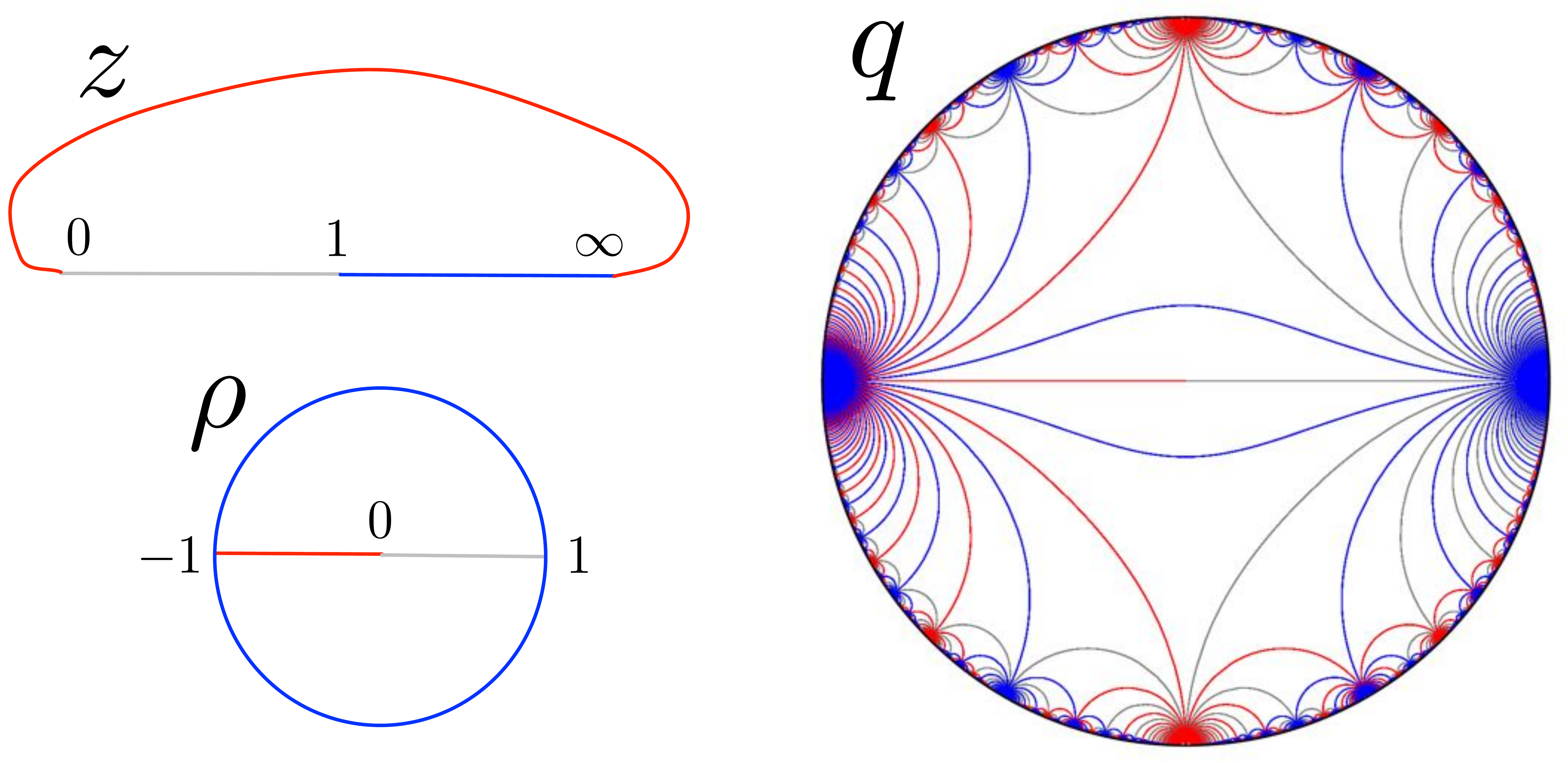}	
\caption{The $q(z)$ map takes the universal cover of the $z$-plane (the sphere with punctures at $0,1,\infty$) to $|q|<1$.  This figure suggests the relationship between the $z$ plane, the unit $\rho$ disk, and the unit $q$ disk, with branch cuts indicated with colored lines \cite{Maldacena:2015iua}. The relations between these variables are $q=e^{-\pi\frac{K(1-z)}{K(z)}} $ and  $z = \frac{4 \rho}{(1+\rho)^2}$, and the inverse transformations are $z=\left(\frac{\theta_2(q)}{\theta_3(q)}\right)^4$ and $\rho=\frac{z}{(1+\sqrt{1-z})^2}$. The Virasoro blocks converge throughout $|q| < 1$, with OPE limits occurring on the $q$ unit circle.}
\label{fig:zrhoqBranchCuts}
\end{figure}

There are two versions of the Zamolodchikov recursion relations (for a nice review see \cite{Perlmutter:2015iya}).  The first \cite{ZamolodchikovRecursion} is based on writing $\CV_{h_i, h,c}$ as a sum over poles in the central charge $c$, plus a remainder term that survives when $c \to \infty$ with operator dimensions fixed.  The second \cite{Zamolodchikovq}, which is more powerful, arises from expanding the blocks as a sum of poles in the intermediate dimension $h$ plus a remainder term that survives as $h \to \infty$.  The remainder term can be computed from the large $h$ limit of the Virasoro blocks \cite{Zamolodchikov:1986gh, Zamolodchikovq}.  This large $h$ limit of the blocks takes a simple form when written in terms of the uniformizing variable
\be\label{eq:qz}
q(z) = e^{i \pi \tau(z)} \equiv e^{- \pi \frac{K(1-z)}{K(z)} }
\ee
where $K(z)$ is the elliptic function
\be
K(z) = \frac{1}{2} \int_0^1 \frac{dt}{\sqrt{ t(1-t)(1-zt) } }
\ee
The $q$-coordinate can be derived from the accessory parameter/monodromy method in the semiclassical limit \cite{HarlowLiouville} or from a quantization of the theory on the pillow metric \cite{Maldacena:2015iua}.  It has  the remarkable feature that $q(z)$ covers the full multisheeted $z$-plane (the sphere with punctures at $0,1,\infty$), as depicted in figure \ref{fig:zrhoqBranchCuts}. The Virasoro blocks can then be written in the form
\be\label{eq:FullVBlock}
\mathcal{V}_{h,h_{i},c}\left(z\right)=\left(16q\right)^{h-\frac{c-1}{24}}z^{\frac{c-1}{24}-h_{1}-h_{2}}\left(1-z\right)^{\frac{c-1}{24}-h_{2}-h_{3}}\left[\theta_{3}\left(q\right)\right]^{\frac{c-1}{2}-4\sum_{i=1}^{4}h_{i}}H\left(c,h,h_{i},q\right)
\nn \\
\ee
 where $H\left(c,h,h_{i},q\right)$ can be obtained from the recursion relation:
\be
H\left(c,h,h_{i},q\right)=1+\sum_{m,n=1}^\infty\frac{q^{mn}R_{m,n} }{h-h_{m,n}}H\left(c,h_{m,n}+mn,h_{i},q\right)
\ee
We note that this recursion relation naturally produces a series expansion in the variable $q$.  For more details along with the definitions of the quantities appearing in these equations see appendix  \ref{app:ZRecursion}.  

In this work, we will be using the recursion relations to obtain the $q$-expansion of the Virasoro blocks to very high orders.  
It appears that most prior implementations of the Zamolodchikov recursion relations could not reach the $N \sim 1000$ that we will study.\footnote{Prior implementations such as this code \cite{HartmanLargeC} and other similar, modestly improved versions we are aware of. Perhaps \cite{Collier:2017shs} are using roughly the same algorithm we describe. We have only used laptops; one could perhaps achieve $N \sim 10^4$ with more computing power.}  Our improvements are fairly elementary, and are based on computing and storing the specific coefficients of powers of $q$ in $H(c,h_{m,n}+mn,h_i,q)$, as we describe in more detail in appendix  \ref{app:ZRecursion}. The computational time complexity of our algorithm is roughly $O(N^3(\log N)^2)$, while it seems that some earlier implementations scaled exponentially with $N$. The maximum $N$ is limited by memory consumption, with memory usage scaling roughly as $O\left(N^3 \log N\right)$.  We have verified our code by comparing to a number of previous results, including prior implementations, the semiclassical blocks, blocks computed by brute force from the Virasoro algebra, and the special case of degenerate external operators.

\subsection{Kinematics and Convergence of the $q$-Expansion}
\label{sec:Kinematicsandq}

Both the correlator and the Virasoro blocks in equation (\ref{eq:GenericVirasoroDefinition}) can have singularities in the OPE limits, which occur when $z \to 0, 1, \infty$.  Generically we expect branch cuts in the $z$-plane running between these three singularities.  So for our purposes, the most remarkable feature of the variable $q(z)$ is that the region $|q| < 1$ covers not only the complex $z$-plane, but also every sheet of its cover. The relationship of the $z$ plane and its branch cuts to the region $|q| < 1$  \cite{Maldacena:2015iua} is depicted in figure \ref{fig:zrhoqBranchCuts}.  The Zamolodchikov recursion relations provide an expansion for the Virasoro blocks that converges for all $|q| < 1$, which means that they can provide a good approximation to the 4-pt correlator in any kinematic configuration.  In particular, we can use the $q$-expansion to study the Lorentzian regime with arbitrary time-orderings for the operators.

The existence of the $q$-variable implies that in CFT$_2$, there are an infinite number of distinct regimes where the bootstrap equation may be analytically tractable.  In the case of $d \geq 3$, one can study the OPE limit $z \to 1$ using conformal blocks expanded in the OPE limit of small $z$, and this implies various exact results about the properties of large spin operators \cite{Fitzpatrick:2012yx, KomargodskiZhiboedov}.  However, because the Euclidean OPE in $d \geq 3$ does not converge deep in the Lorentzian region, one cannot study other OPE channels in the same way.  This obstruction disappears in $d =2$, where one must be able to reproduce all of the distinct OPE limits $|q| \to 1$ pictured in figure \ref{fig:zrhoqBranchCuts} using the small $q$ expansion.   The large Lorentzian time limit that we will discuss in section \ref{sec:LateTime} provides a physically motivated example of this idea.

\begin{figure}
\centering
\includegraphics[width=0.4\textwidth]{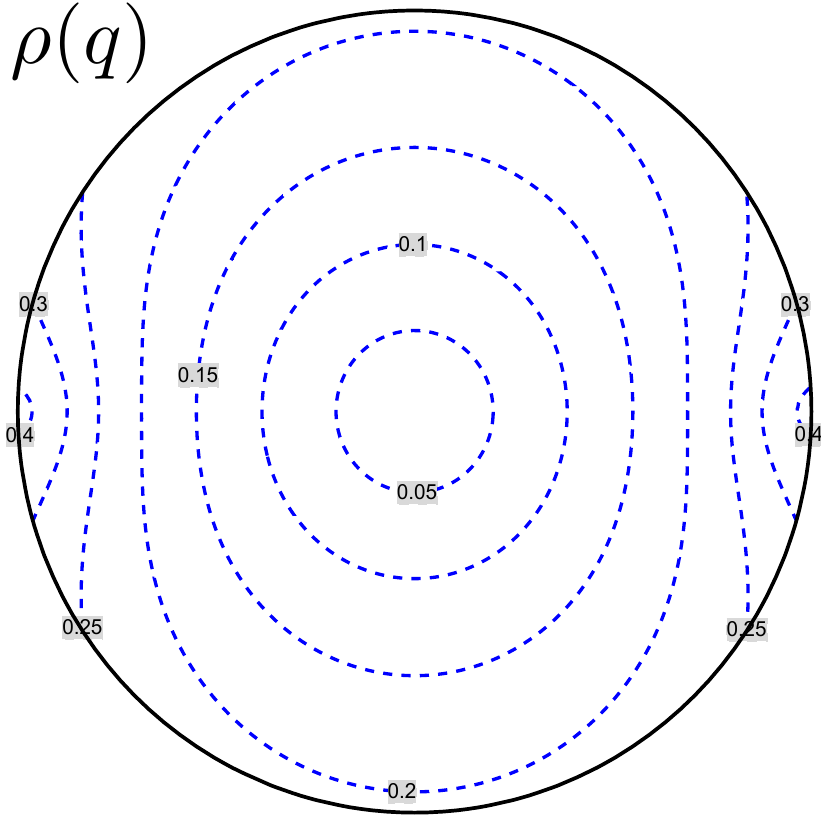}	
\caption{This figure displays contours of constant $|q|$ inside the $\rho$ unit circle, which corresponds to the entire $z$-plane  via $z = \frac{4 \rho}{(1+\rho)^2}$. Since this is only the first sheet of the $z$-plane, it corresponds to the region in the $q$-disk enclosed by the two blue lines connecting $\pm1$ in figure \ref{fig:zrhoqBranchCuts}.  The correlator can have singularities in the OPE limits $\rho \to -1, 1$ and these correspond to $q \to -1, 1$ as well.  Away from these limits $|q| < |\rho |$ and the $q$-expansion converges much more rapidly than the $\rho$ expansion.}
\label{fig:Constantq}
\end{figure}

\begin{figure}
\centering
\includegraphics[width=0.85\textwidth]{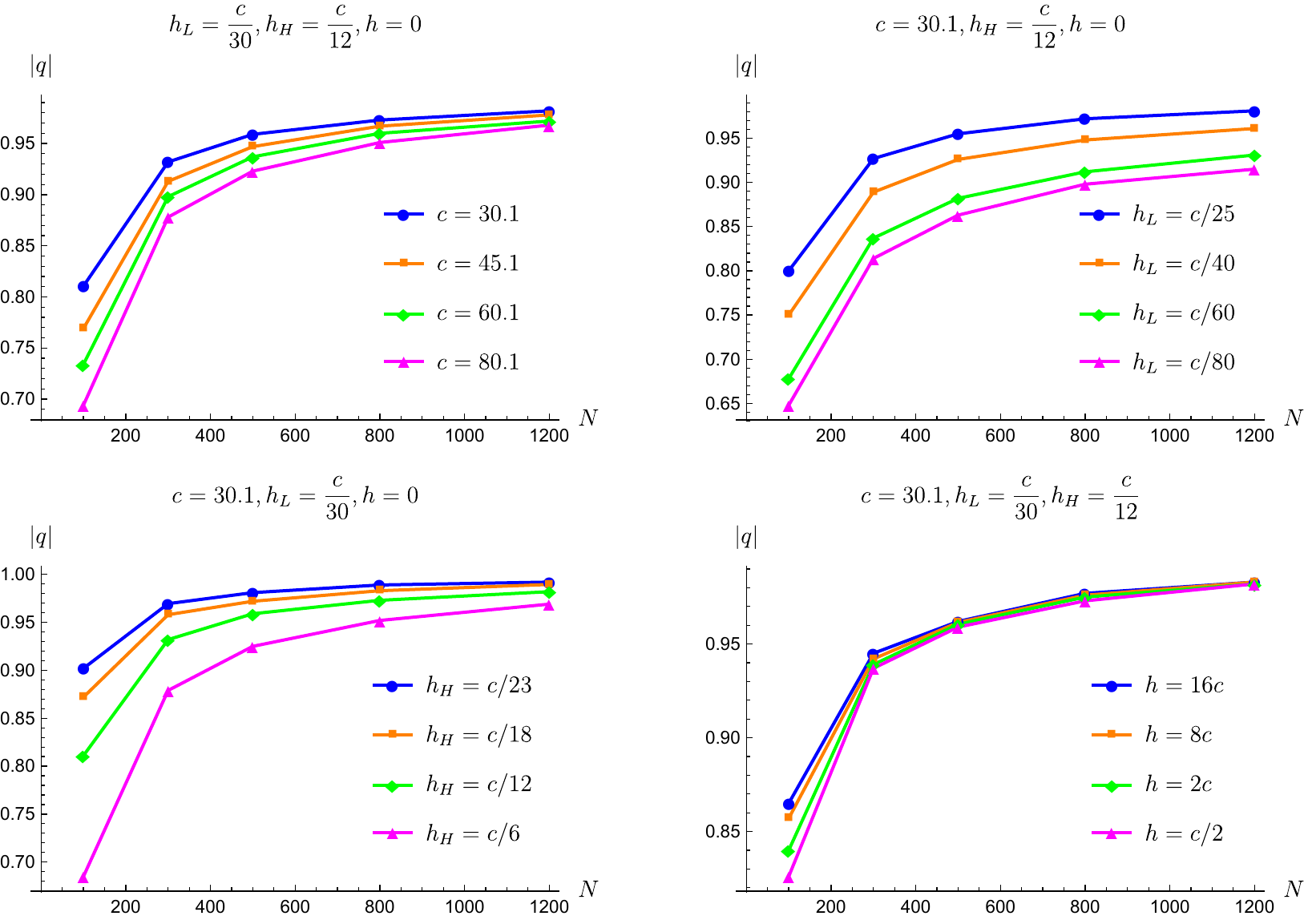}	
\caption{These plots display the maximum  $|q|$ where the $q$-expansion converges for various choices of parameters.  Convergence improves when $h_L$ and $h_H$ move closer to $c/24$ and when $c$ decreases.  The intermediate primary dimension $h$ seems to have little effect on convergence.   These plots define `convergence' as $\left|\left|\frac{\CV_{0.95N}(q)}{\CV_N(q)}\right|-1\right|<10^{-5}$, where $\CV_M$ includes an expansion up to order $q^M$.} 
\label{qConvergentRange}
\end{figure}

We will be studying numerical approximations to the Virasoro blocks based on a large-order expansion in the $q$ variable.  Thus to understand the convergence properties of our expansion, it may be useful to map out the regions of constant $|q|$.  For this purpose we can use the coordinate $\rho(z)$ defined via $z = \frac{4 \rho}{(1+\rho)^2}$ \cite{Pappadopulo:2012jk}, because the entire $z$-plane can be easily visualized as the region $|\rho| < 1$.  The operators at $z= 1$ and $\infty$ are mapped to $\rho = 1$ and  $-1$, respectively.  In figure \ref{fig:Constantq} we have plotted contours of constant $|q|$ in the $\rho$-coordinate system.  In figure \ref{qConvergentRange} we present results on the convergence region of the $q$-expansion of the Virasoro blocks for various values of the dimensions and central charge.

\begin{figure}
\centering
\includegraphics[width=0.65\textwidth]{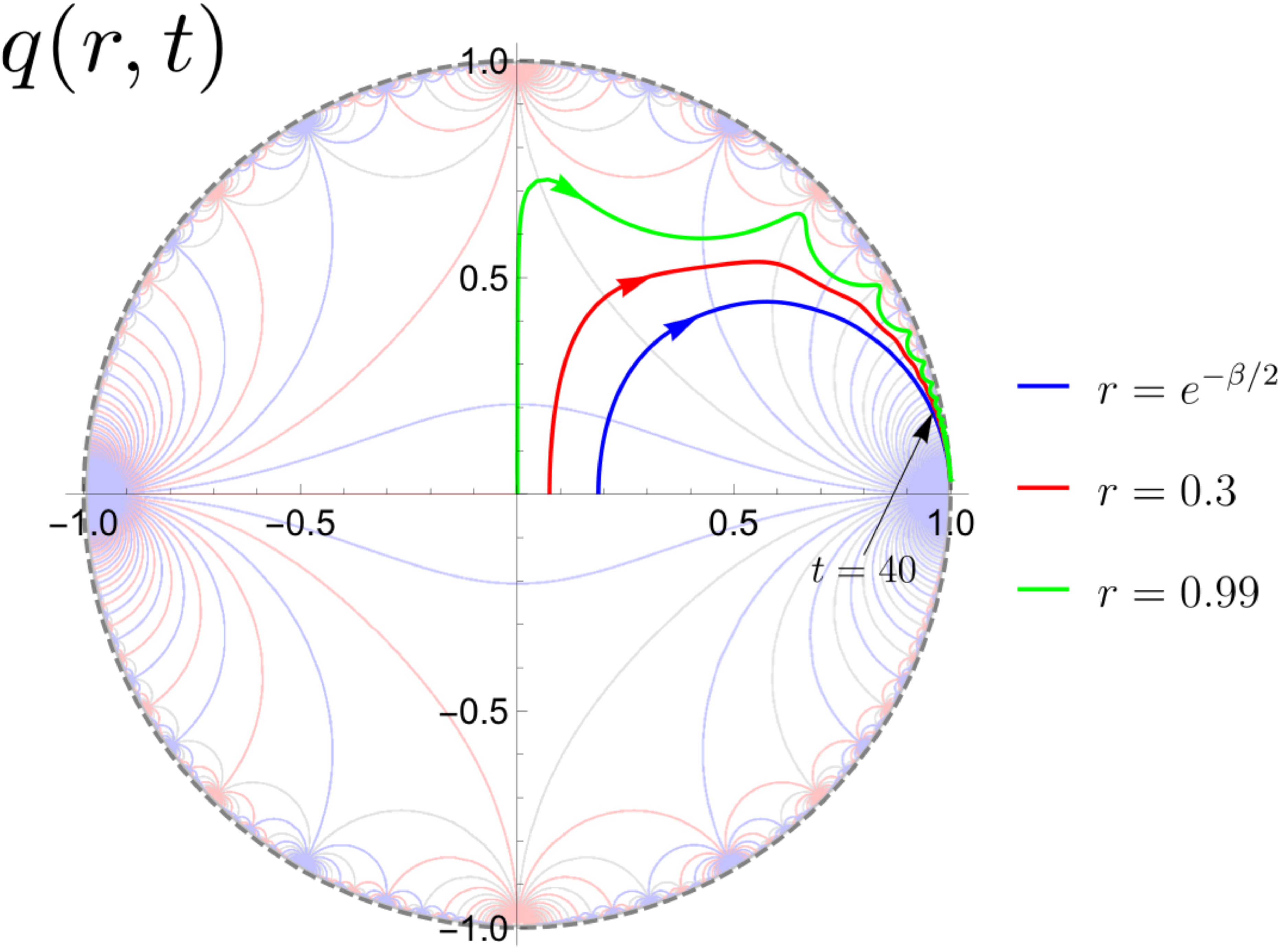}	
\caption{The function $q(r,t)$ for different $r$, where we have written $z = 1 - r e^{-it}$ and plotted the lines $t \in [0, \infty)$. For the blue curve we chose $\beta = 2 \pi$, which corresponds with $h_H = \frac{c}{12}$.  Note that the wiggles are due to the fact that when $t = 2 \pi n$ and $r \approx 1$, the coordinate $z$ approaches an OPE singularity. The large time limit of $q$ was given in equation (\ref{eq:LargeTimeqApprox}).}
\label{fig:qoft}
\end{figure}

A kinematic configuration that will be of particular interest represents $z = 1 - r e^{-i t}$  (and $\bar z = 1 - r e^{-it}$ as well) and is  depicted in the AdS/CFT context in figure \ref{fig:TimelikeSeparation}.   With this setup we can study the correlator of light operators $\CO_L(z) \CO_L(0)$ at timelike separation in the background created by a heavy operators $\CO_H$.  At large times $t$, this correlator can be used as a probe of information loss in pure state black hole backgrounds, as we will discuss in section \ref{sec:LateTime}.  On the $z$ plane, the late time behavior is obtained by analytically continuing the conformal block around the branch-cut starting at $z=1$ multiple times. Explicitly, the Lorentzian value of the $q$ variable is obtained with the following analytic continuation of the elliptic integral: 
\begin{equation}
\left.K\right|_{z\rightarrow1-e^{-it}}=K(1-re^{-it})-2i \left\lfloor \frac{1}{2}-\frac{t}{2\pi}\right\rfloor K(e^{-it}),
\end{equation}
where the elliptic functions on the right-hand side are evaluated on the principle sheet with the branch-cut chosen to be $z\in[1,\infty)$. At large $t$ we have
\be
q(t) \approx 1 + \frac{i \pi^2}{t} - \frac{\pi^4 + 2\pi^3 g(r,t)}{2 t^2} + \cdots 
\label{eq:LargeTimeqApprox}
\ee
where
\be
g(r,t) = \frac{K\left(1-e^{-i t} r\right)}{K\left(e^{-i t} r \right)}+2 i \left\lfloor \frac{t+\pi}{2 \pi }\right\rfloor -\frac{i t}{\pi } 
\ee
with the elliptic function $K(z)$ are taken on their principle sheet, so that $g(r,t)$ is periodic in $t$.    This means that $|q|^2 \approx 1 - \frac{\pi^3}{t^2} (g + g^*) + \cdots$ and the real part Re$[g(r,t)] > 0$, so that $|q| < 1$ for all $t$, as expected.  In the limit that $r \ll 1$, we have $g(r,t) \approx \frac{1}{\pi} \log \frac{16}{r}$, which leads to the estimate  
\be
|q|^2 \approx 1 - \frac{2 \pi^2 \log \frac{16}{r} }{t^2}
\ee
in the limit of $r \ll 1$ and $t \to \infty$.  Thus we can translate between convergence in $|q|$ and $t$; very roughly, we expect that working to order $q^N$ will allow us to probe $t \propto \sqrt{N}$ at large $N$.  We can visualize the trajectory of $q(r,t)$ for various $r$ in figure \ref{fig:qoft}.

\subsection{Review of Blocks and Adherence to the Semiclassical Form}
\label{sec:AdherencetoSemiclassicalLimit}

Much is known about the Virasoro blocks in various limits.    In the limit $c \to \infty$ with all dimensions held fixed, the Virasoro blocks simply reduce to global conformal blocks, which are hypergeometric functions.  Corrections to this result up to order $1/c^3$ are known explicitly \cite{Chen:2016cms}.  In the heavy-light limit, where we take $c \to \infty$ with two `heavy' operator dimensions $h_H \propto c$, and the two `light' dimensions $h_L$ and the intermediate operator dimension $h$ fixed, the blocks  take the form \cite{Fitzpatrick:2015zha}  
\be \label{eq:HeavyLightBlocks}
\CV = (1-w)^{h_L \frac{\alpha_H - 1}{\alpha_H}} \left( \frac{w}{\alpha_H} \right)^{h-2h_L} {}_2 F_1(h,h,2h, w)
\ee
where $w \equiv 1 - (1-z)^{\alpha_H}$ and $\alpha_H \equiv \sqrt{1 - \frac{24 h_H}{c}}$.  Note that when $h_H > \frac{c}{24}$, we have $\alpha_H = 2 \pi i T_H$ where $T_H$ is the Hawking temperature of a corresponding BTZ black hole.
In the case of the vacuum block, which is $h=0$, the $1/c$ corrections to this limit are also known explicitly \cite{Fitzpatrick:2015dlt} for any $h_H/c$.  Finally, in the semiclassical large $c$ limit, where all dimensions $h_i, h\propto c$, there is overwhelming evidence that the blocks take the form
\be
\label{eq:SemiclassicalLimitofBlocks}
\CV = e^{-\frac{c}{6} f \left( \frac{h_i}{c}, \frac{h}{c}, z \right)} 
\ee
as though they are derived from a semiclassical path integral (and in fact  they have an sl$(2)$ Chern-Simons path integral representation \cite{Fitzpatrick:2016mtp}).  The semiclassical saddle points  have been classified \cite{Fitzpatrick:2016mjq}, and in some kinematic limits we can determine the behavior of $f$ analytically.  In particular, the large Lorentzian time behavior of $f$ with the kinematics of figure \ref{fig:TimelikeSeparation} and  $h_L < \frac{c}{24} < h_H$ has been determined  \cite{Fitzpatrick:2016mjq}.  The result is that the leading semiclassical contribution always decay exponentially at sufficiently large times\footnote{As we increase the intermediate operator dimension this behavior may not set in until later and later times.  Here we are studying late times with all other parameters held fixed.} at the rate 
\be \label{eq:SemiclassicalLateTimeDecay}
\CV(t) \approx \exp \left[ -\frac{c}{12}    | \alpha_H |  \left(1 - \alpha_L \right) |t| \right]
\ee
where $\alpha_L = \sqrt{1 - \frac{24 h_L}{c}}$ and $\alpha_H = 2 \pi i T_H$ with $T_H$ the corresponding Hawking temperature.  As we will review in section \ref{sec:LateTime}, this demonstrates that information loss due to black hole physics \cite{Maldacena:2001kr} occurs as a consequence of the behavior of the individual Virasoro blocks  \cite{Fitzpatrick:2016ive, Fitzpatrick:2016mjq}.  Finally, some exact information about the behavior of the Virasoro blocks can be obtained by studying degenerate states \cite{Fitzpatrick:2016ive}.

Most of these approximations hold in the large central charge limit when the kinematic configuration is held fixed.  But the deviations between the exact and semiclassical Virasoro blocks may depend importantly on the kinematics.  As we will discuss in detail below, the semiclassical blocks have `forbidden singularities' that are absent from the exact blocks \cite{Fitzpatrick:2016ive}.  We also find that as expected \cite{Fitzpatrick:2016ive, Fitzpatrick:2016mjq}, the exact and semiclassical blocks have very different behavior at large Lorentzian times.  More generally, we would like to map out the kinematic regimes where non-perturbative corrections to the semiclassical Virasoro blocks become large.  

\begin{figure}
\centering
\includegraphics[scale=0.95]{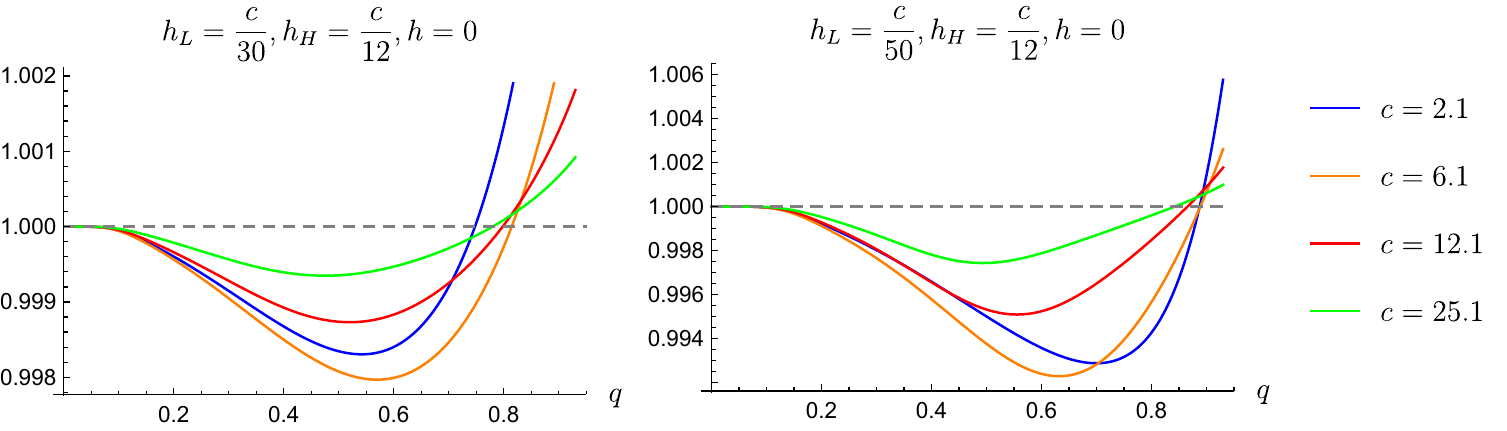}	
\caption{In this figure we plot $R = \frac{ c' \log \CV(c,q)}{c \log \CV(c',q)}$ with  $c' \equiv \frac{11}{10}c$ in order to test the semiclassical limit. As we increase $c$, the semiclassical limit becomes a better approximation and $R \to 1$, but even for $c = 2.1$ the blocks are remarkably well approximated by the semiclassical form. For the larger choices of $c$ the functions have similar shapes up to an overall rescaling; this suggests that the first $1/c$ correction is dominating the discrepancy $R-1$.  In the OPE limit $q\rightarrow 0$ the semiclassical limit always applies. We find similar results for non-vacuum blocks. }
\label{fig:RatioOfLogV}
\end{figure}

But at a more basic level, it is interesting to ask how large $c$ must be before the semiclassical limit of the Virasoro blocks provides a reasonable approximation to their behavior.  This has immediate implications for the possibility of constructing a 2d CFT and probing quantum gravity in an experimental lab.  A natural way to probe the existence of the semiclassical limit is by studying the ratio of logarithms of blocks
\be
R \equiv \frac{ c_2 \log \CV(c_1,q)}{c_1 \log \CV(c_2,q)} \stackrel{?}{=} 1
\ee
at somewhat different central charges $c_1$ and $c_2$.  If the semiclassical limit of equation (\ref{eq:SemiclassicalLimitofBlocks}) is a good approximation, then this quantity will be $1$, but otherwise we expect it to deviate from $1$ by effects of order $\frac{1}{c}$. In figure \ref{fig:RatioOfLogV} we explore this ratio and find that the semiclassical form $\CV \approx e^{c f}$ provides a remarkably good approximation for very small values of $c$.

There is an important caveat that we will return to in section \ref{sec:ForbiddenSingularitiesandReconstruction}.  An infinite number of distinct semiclassical saddle points can contribute to the Virasoro blocks in the large $c$ limit \cite{Fitzpatrick:2016mjq}.  Thus it is possible that $\CV \approx e^{c f}$ for some $f$, but that due to Stokes phenomena, the dominant saddle $f$ changes as we move in the $q$ unit disk.  So although the semiclassical limit may appear to describe the blocks well for all $q$, as indicated by figure \ref{fig:RatioOfLogV}, in fact the saddle that is leading near $q \approx 0$ may be sub-leading at general $q$.  Thus  the naive semiclassical blocks may differ greatly from the exact blocks; in fact we will find this to be the case in section \ref{sec:ForbiddenSingularitiesandReconstruction}.  

Nevertheless, figure \ref{fig:RatioOfLogV} suggests that we should be optimistic about probing semiclassical CFT$_2$ correlators in the lab!  It would be very interesting to engineer a CFT$_2$  with $c > 1$  and no conserved currents aside from the stress tensor \cite{Plamadeala:2014roa}.

\section{Late Time Behavior and Information Loss}
\label{sec:LateTime}

One sharp signature of information loss in AdS/CFT \cite{Maldacena:2001kr} is the exponential decay of correlation functions at large time separations in a black hole background.  This can be studied using heavy-light 4-point functions in CFT \cite{Fitzpatrick:2014vua}.  As portrayed in figure \ref{fig:TimelikeSeparation}, we can interpret this correlator as the creation and subsequent measurement of a small perturbation to an initial high-energy state.  In a unitary theory on a compact space with a finite number of local degrees of freedom, this initial perturbation cannot completely disappear.  But a computation in the black hole background displays eternal exponential decay, capturing the physical effect of the signal falling into the black hole.  At a more technical level, the exponential decay rate can be obtained from the quasinormal mode spectrum of fields propagating in the black hole geometry.

The simplest way to see that heavy-light correlators cannot decay forever is to expand in the $\CO_H \CO_L \to \CO_H \CO_L$ channel, giving
\be \label{eq:EnergyDecomp}
\< \CO_H(\infty) \CO_L(t) \CO_L(0) \CO_H(-\infty) \> = \sum_E   \lambda^2(E) e^{i E t}
\ee
where $\lambda^2(E)$ is a product of OPE coefficients.  Because the sum on the right-hand side is discrete, the correlator must have a finite average absolute value at late times.
When $h_{H} \gtrsim \frac{c}{24} \gg h_{L}$, we expect the states contributing in (\ref{eq:EnergyDecomp}) to be a chaotic collection of $e^{S}$ blackhole microstates with energy near that of $\CO_{H}$, and with $S = \frac{\pi^2}{3} T_H c$. The amplitude will initially decay  due to cancellations between the essentially random phases, but  these cancellations cannot become arbitrarily precise.  Roughly speaking, the decay should stop when the correlator reaches $\sim e^{-S}$ and begins to oscillate chaotically.  At a more detailed level, the time dependence can change qualitatively on timescales of order $S$ and $e^S$ as different features of $\lambda^2(E)$ come into play \cite{Barbon:2014rma, Cotler:2016fpe, Dyer:2016pou}.

In this work, we will not study the $\CO_H \CO_L \to \CO_H \CO_L$ channel directly.   Instead we work in the channel where $\CO_{H} \CO_{H} \to \CO_{L} \CO_{L}$, which is related to the first channel by the bootstrap equation (or by modular invariance in the case of the partition function \cite{Dyer:2016pou}).  In this channel we are sensitive to the exchange of states between the heavy and light operators.  For example, pure `graviton' states in AdS$_3$ correspond to the exchange of the Virasoro descendants of the vacuum, which are encapsulated by the Virasoro  vacuum block.  Other heavy-light Virasoro blocks include a specific primary state along with its Virasoro descendants, which one can think of as gravitational dressing.  We are interested in this channel because heavy-light Virasoro blocks encode many of the most interesting features of semiclassical gravity.  We would like to understand to what extent the exact Virasoro blocks know about the resolution of information loss.

It is convenient to think of the time dependence of the Virasoro blocks as coming from a potentially continuous $\lambda_h(E)$ associated with each block, via
\be \label{eq:VasEIntegral}
\CV_h(t) = \int dE \, \lambda^2_h(E) e^{i E t}
\ee
where $h$ labels the dimension of an intermediate Virasoro primary operator $\CO_h$  in both the $\CO_{H}(x) \CO_{H}(0)$ and  $\CO_{L}(x) \CO_{L}(0)$ OPEs.  Roughly speaking, the late time dependence of $\CV_h(t)$ will come from the least analytic features of $\lambda_h^2(E)$. 

For example, in the leading semiclassical limit, heavy-light Virasoro blocks decay exponentially at late times at a universal rate given in equation (\ref{eq:SemiclassicalLateTimeDecay}).   This semiclassical behavior comes from a function $\lambda_h^2(E)$ that is smooth on the real axis, but has poles in the complex $E$-plane. In AdS$_3$ these poles can be interpreted as the quasinormal modes of a BTZ black hole background (at least for small $h$).  A straightforward contour deformation of equation (\ref{eq:VasEIntegral}) connects these poles to the exponential decay.

At sufficiently late times, the physics of the quasinormal modes will be subdominant to less analytic features in $\lambda_h^2(E)$.  For example, if $\lambda_h^2(E)$ exhibits thresholds of the form  $(E-E_*)^{p-1}$ with $E_*$ real, then $\CV(t)$ will inherit a power-law behavior $t^{-p}$ at late times.  And if $\lambda_h^2(E)$ receives delta function type contributions, then $\CV(t)$ will have a finite average absolute value at late times.  If such features are present in $\CV_h(t)$, then it is natural to investigate the timescale where $\CV_h(t)$ transitions from exponential decay to some other late-time behavior.  

The full CFT$_2$ correlator should not become much smaller than $\sim e^{-S}$.  Since Virasoro blocks associated with light operators initially decay exponentially, one might naively expect that $\CV_h(t)$ should change qualitatively after a time of order $S$.  More specifically, for heavy-light correlators dominated by the vacuum block, we would expect a departure from exponential decay by a time
\be\label{eq:deftD}
t_D = \frac{\pi c}{6 h_L}
\ee
up to an unknown order one factor.   This argument is rather weak, since the full correlator might not behave like the light-operator Virasoro blocks.  However, the same prediction for $t_D$ was derived from an analysis of non-perturbative effects \cite{Fitzpatrick:2016ive} in the vacuum block.  We discuss the equation that led to that prediction in section \ref{sec:FateSemiclassical}.

We will see empirically that Virasoro blocks with small $h$ do undergo a transition at a timescale remarkably close to $t_D$.  Furthermore, at late times the behavior of the heavy-light Virasoro blocks appears to be a universal power-law:
\begin{equation}
|\mathcal{V}_{h_{L},h_{H},h,c}\left(t\gg t_D\right)| \propto t^{-\frac{3}{2}},
\end{equation}
where we require $h_H\ge\frac{1}{24}$, so that at least one external operator is heavy enough to create a blackhole.  When the intermediate dimension $h \gtrsim h_H$ the late time power-law behavior remains the same, although the transition time then also depends on $h$ (and we do not have an analytic prediction to compare to).  This universal behavior suggests a threshold $\sqrt{E - E_*}$ in $\lambda_h^2(E)$, which seems to correspond with random matrix behavior \cite{Guhr:1997ve, Cotler:2016fpe}.  Our results indicate that  the $t^{-\frac{3}{2}}$ power-law persists to timescales $\sim e^S$, so individual heavy-light Virasoro blocks are not sensitive to the discreteness of the spectrum.

These results show that the time-dependence of the heavy-light Virasoro blocks has some qualitative similarities with that of the Virasoro vacuum character after an $S$ transformation and the analytic continuation $\beta \to \beta + i t$ \cite{Dyer:2016pou}.  Both the heavy-light blocks with small $h$ and the vacuum character have an initial exponential-type decay, though the precise time-dependence is rather different.  The heavy-light blocks and the vacuum character have the same power-law decay at late times, though non-vacuum characters decay with a different late-time power-law \cite{Dyer:2016pou}.

In what follows we will study the heavy-light blocks $\CV_h(t)$ empirically to establish the robust features of their time-dependence.  We also translate the late-time $t^{-3/2}$ behavior into a statement about the coefficients of $q^N$ in $\CV_h(q)$ at large orders in the $q$-expansion, as one might hope to derive this asymptotic behavior for the coefficients using the Zamolodchikov recursion relations.  One might also compute $\lambda_h^2(E)$ directly using the crossing relation \cite{Ponsot:1999uf, Teschner:2003en}. Finally  we discuss the implication of our results for the  late time behavior of the correlator.

\begin{figure}[h]
\begin{centering}
\includegraphics[width=0.98\textwidth]{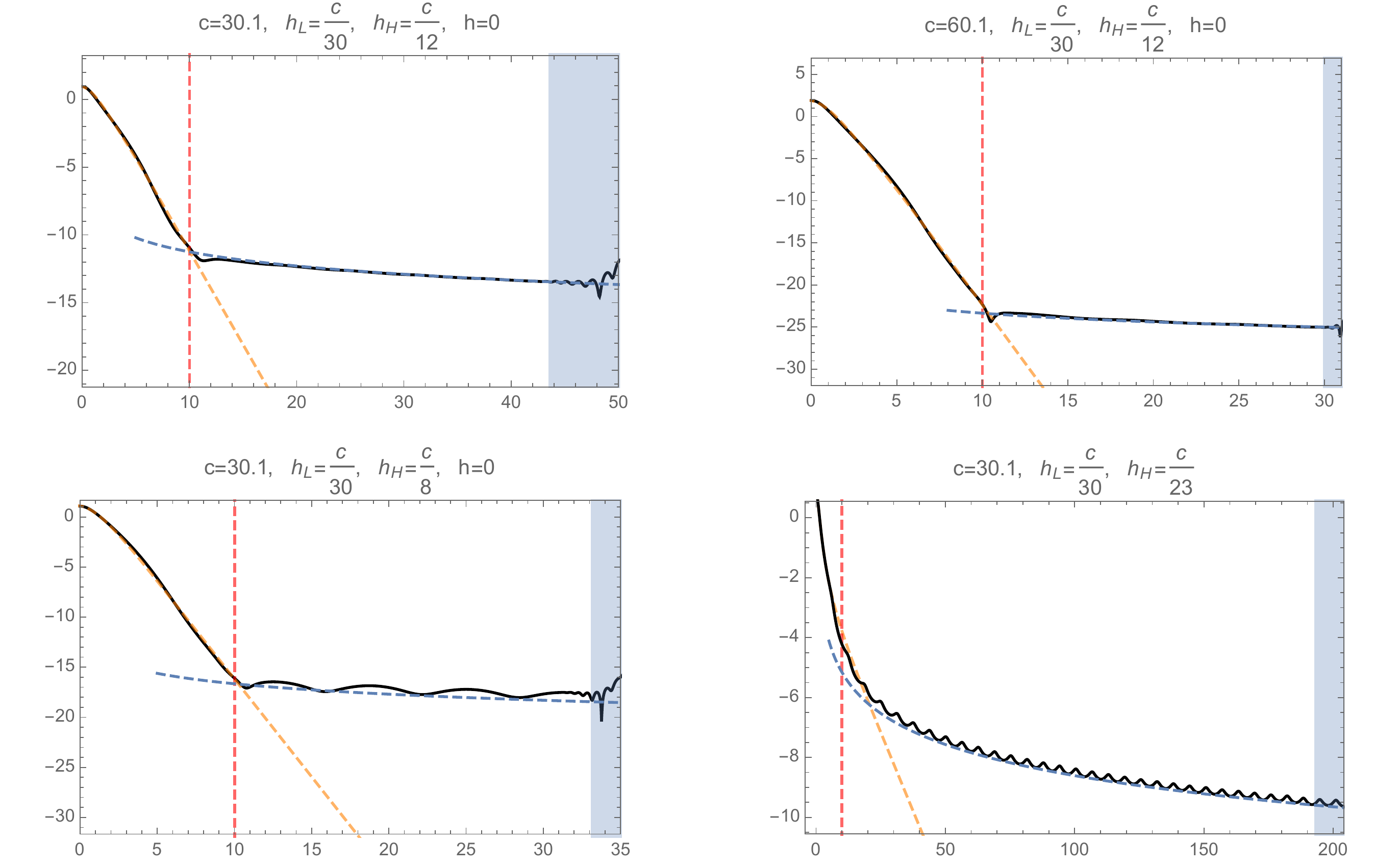}
\caption{Heavy-light Virasoro vacuum blocks switch from an initial exponential decay to a slow, universal power law decay at roughly the time scale $t_d = t_D - b$, where the constant offset $b$ depends on the choice of $r$ in $z = 1 - re^{-it}$. The vertical axis is $\log|\mathcal{V}|$, while the horizontal axis is the Lorentzian time $t$. The black lines are full Virasoro vacuum blocks computed to order $q^{1200}$. This polynomial truncation stops converging in the shaded region. The yellow dashed lines are the semiclassical vacuum blocks using methods of \cite{Fitzpatrick:2016mjq}. The red dashed lines are the time scale (\ref{eq:deftD}). The blue dashed lines are the power law $a t^{-\frac{3}{2}}$ with $a$ properly chosen to match the full blocks.  }
\label{fig:VBlock}
\end{centering}
\end{figure}

\subsection{Numerical Results and Empirical Findings}

\subsubsection{Vacuum Virasoro Blocks}

Using the methods discussed in section \ref{sec:BlocksZRR}, we compute the vacuum Virasoro blocks at late times. Figure \ref{fig:VBlock} shows the result along with a comparison to the semiclassical blocks computed using semi-analytic methods \cite{Fitzpatrick:2016mjq}.   For numerical convenience we avoid certain rational values of $c$ to prevent singularities in intermediate steps of the computation. 

\begin{figure}[h]
\begin{centering}
\includegraphics[width=0.65\textwidth]{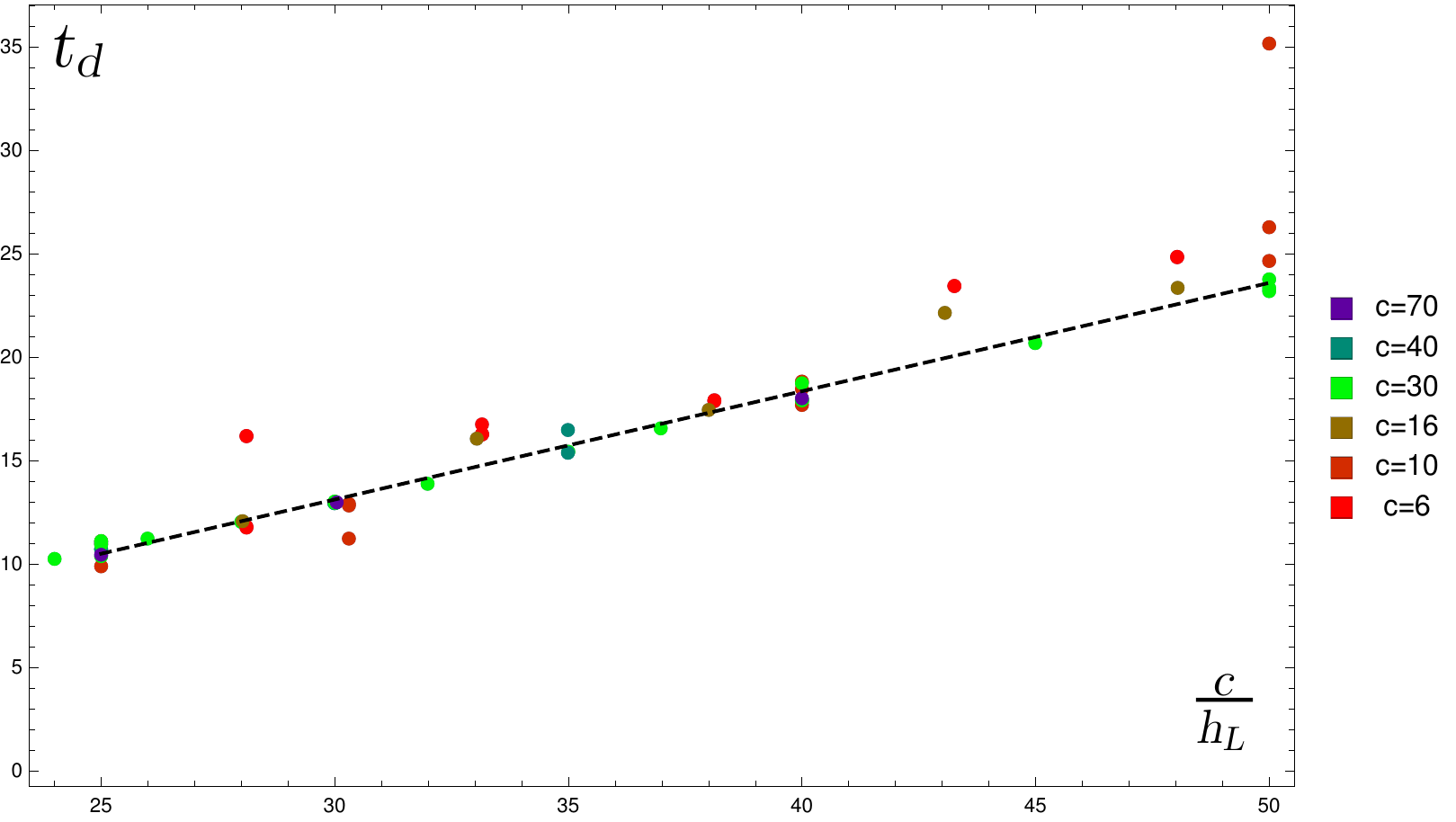}
\caption{This figure displays the time $t_d$ at which the semiclassical vacuum blocks drop below the exact vacuum blocks.  The dashed line is a fit to the analytic prediction $t_D \equiv \frac{\pi c}{6 h_L}$ with an empirical offset $t_d = t_D -  2.6$; the offset depends on the choice of $r$ with $z = 1 - r e^{-it}$. Note that the data with smaller values of $c$ is noisy, but the larger values fit the linear behavior extremely well.  The plot includes a variety of choices for $\frac{h_H}{c}$. 
\label{fig:td}}
\par\end{centering}
\end{figure}

Using the numerical result of the full Virasoro blocks, we can measure the departure time $t_d$ when the semiclassical block drops below the exact block. We compare this measured value to the prediction of (\ref{eq:deftD}) in figure \ref{fig:td}.  The logic leading up to  (\ref{eq:deftD}) is only valid parametrically, so it is remarkable that it agrees with the measured $t_d$ up to a small constant shift.  Note that we parameterize the time dependence via $z = 1 - r e^{-it}$, and this constant shift depends on $r$.  We have also checked that $t_d$ is primarily controlled by the ratio $\frac{h_L}{c}$ and has a very weak dependence on  $h_H$ and $c$.  

Around the time  $t_D$, all vacuum blocks show an obvious change of behavior from an initial exponential decay to a much slower power law decay.  To very good accuracy, the power of this decay seems to be $t^{-\frac{3}{2}}$ universally in all of the parameter space we were able to explore with an external operator with dimension $h_H>\frac{1}{24}$.  A few examples are provided in figure \ref{fig:VBlock}, but we tested this behavior with hundreds of different parameter choices.

\begin{figure}[h]
\centering{}\includegraphics[scale=0.5]{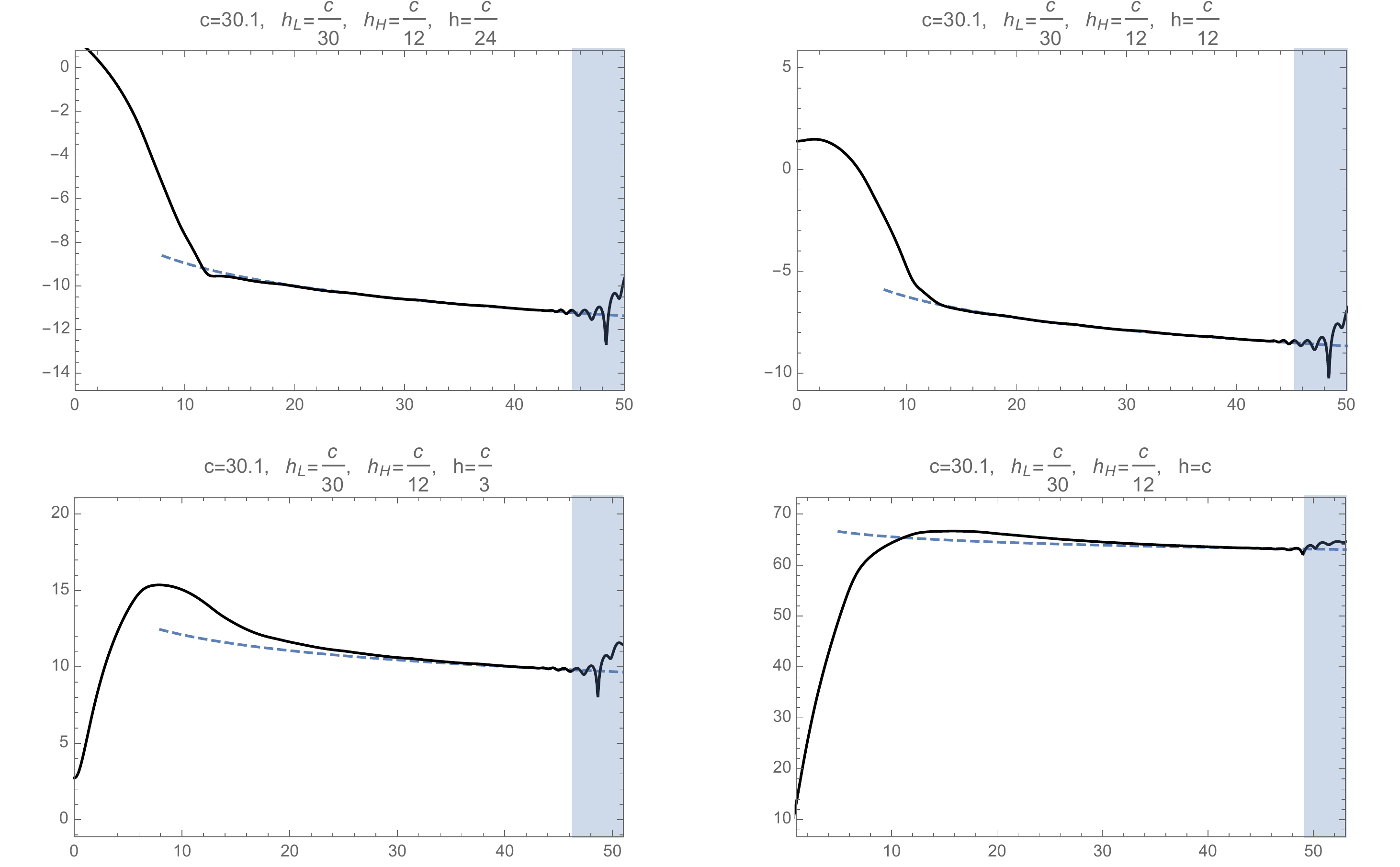}
\caption{The late time behavior of various non-vacuum Virasoro blocks. The vertical axis is $\log|\mathcal{V}|$ and the horizontal axis is the time $t$. The black lines are full Virasoro blocks computed to order $q^{1200}$, plotted using $z = 1 - r e^{-it}$ with $r=0.3$. The polynomial truncation no longer converges  in the shaded region. The blue dashed lines are the power law $a \, t^{-\frac{3}{2}}$ with the constant $a$ fitted to the blocks. We refer to the time and height  of the maxima as $t_{\max}$ and $|\CV|_{\max} = 16^{h - \frac{c-1}{24}} |\tilde \CV|_{\max}$.
\label{NonVBlock}}
\end{figure}

\subsubsection{General Virasoro Blocks}

The non-vacuum blocks also exhibit universal $t^{-\frac{3}{2}}$ late-time decay. The difference from the vacuum case is that we no longer have a simple estimate for the time scale of the transition. In particular, we find that generically the non-vacuum blocks grow at early times, reach a maximum at time $t_{\max}$, and then start to  decay, finally settling down to the $t^{-\frac{3}{2}}$ power law behavior. These features are illustrated by examples in figure \ref{NonVBlock}. 

\begin{figure}[h]
\begin{centering}
\includegraphics[width=0.99\textwidth]{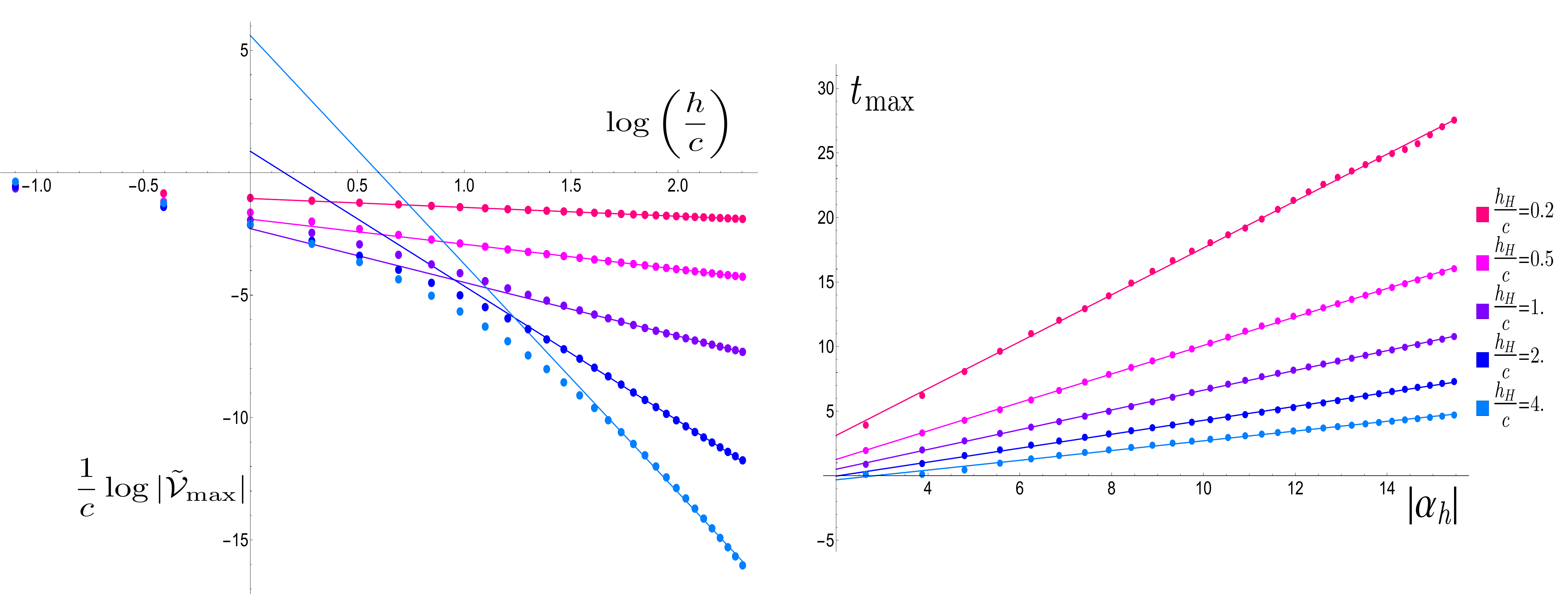}
\caption{These plots show a host of data demonstrating that $|\tilde \CV_{\max}|$ and $t_{\max}$ have simple dependence on $\frac{h}{c}$ when $h \gtrsim h_H$ (recall $\alpha_h \equiv \sqrt{1 - \frac{24h}{c}}$) for a large variety of different choices of $h_H$.  For all of these plots we choose $c=10$, but we have found that the results are robustly $c$-independent.  These plots use $h_L = \frac{c}{30}$, but $h_L$ dependence is mild, as seen in figure \ref{fig:aparameterfits}.  }
\label{fig:LotsofDataforMaxandTime}
\end{centering}
\end{figure}

\begin{figure}[h]
\begin{centering}
\includegraphics[width=0.99\textwidth]{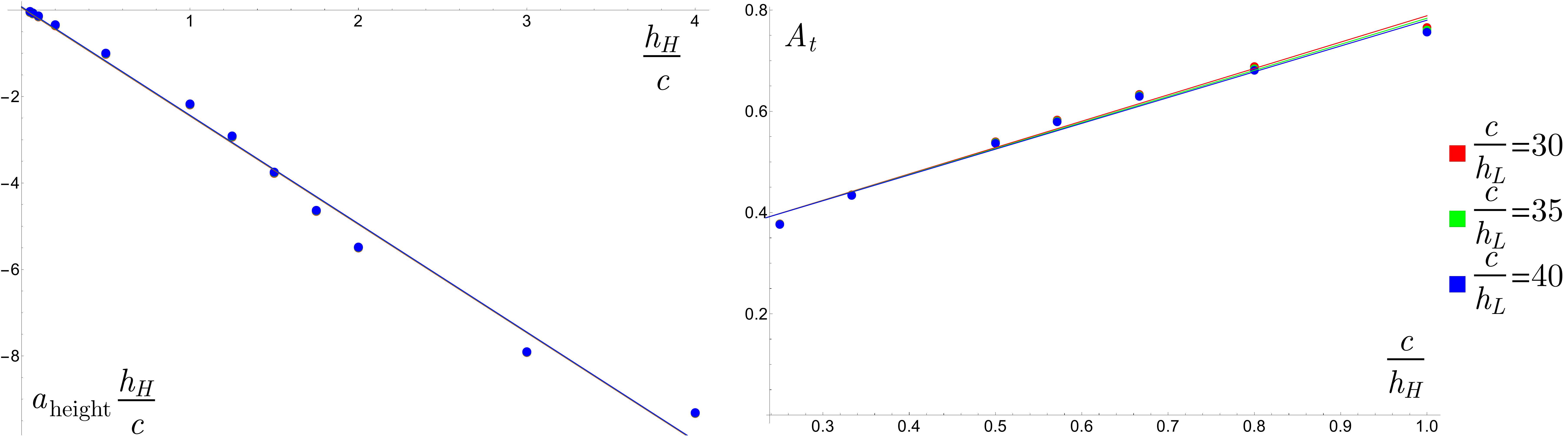}
	\caption{ We have found empirically that the time and height of the maxima of heavy-light Virasoro blocks have a simple dependence on both $h$ and $h_H$.  This figure shows linear fits used to obtain the parameters $a_{\text{height}}$ and $A_t$ defined in equations (\ref{eq:Vmaxdef}) and (\ref{eq:tmaxdef}).  These plots both have $c=10$. Each point is obtained from the slope of $\frac{\log|\tilde{\CV}_\text{max}|}{c}$ and $t_{\max}$ as linear functions of $\log \frac{h}{c}$ and $|\alpha_h|$ respectively (we've used points with $\frac{h}{c} = \frac{n}{3}$ for $n = 1, 2, \cdots, 30$).  We find that both plots are robustly $c$-independent for $c \gtrsim 5$, as expected in the semiclassical limit.  We see explicitly that there is little dependence on $h_L$; in the $a_{\text{height}}$ plot the variation with $h_L$ is almost invisible. }
\label{fig:aparameterfits}
\end{centering}
\end{figure}

From the data plotted in figure \ref{fig:LotsofDataforMaxandTime}, we see that beyond the blackhole threshold $h > \frac{c}{24}$, the timescale $t_{\text{max}}$ has a simple dependence on parameters.  We can fit it to the ansatz 
\be \label{eq:tmaxdef}
t_{\text{max}} = A_t  |\alpha_h| + b_{\text{time}}
\ee
with $\alpha_h=\sqrt{1-\frac{24h}{c}}$ and obtain $A_t$ and $b_{\text{time}}$ empirically.  The parameter $A_t$ is almost a linear function of $\frac{c}{h_H}$, as can be seen in figure \ref{fig:aparameterfits}, with virtually no dependence on other parameters such as $h_L$. It approaches $A_t \approx \frac{c}{2h_H}+\text{constant}$ when $h_H \gtrsim \frac{c}{2}$.  For smaller values of $h_H$ we find  $\frac{1}{2} \geq \frac{d A_t}{d(c/h_H)} \geq \frac{1}{5}$.    We cover a larger range of $h_H$ in figure \ref{fig:at_large_hh} in the appendix, which displays the variation in $A_t$.

On the left of figure \ref{fig:LotsofDataforMaxandTime} we plot $|\tilde \CV_{\max}|$, which is the maximum of the absolute value of the block after extracting a universal prefactor via $|\CV_{\max}| =16^{h - \frac{c-1}{24}} |\tilde \CV_{\max}|$.  We see that $|\tilde \CV_{\max}|$  also has a simple dependence on $\frac{h}{c}$.  We can perform a similar fit for $|\tilde{\CV}_{\max}|$, and we find that
\be \label{eq:Vmaxdef}
\frac{\log |\tilde \CV_{\max}|}{c}  =a_{\text{height}} \frac{h_H}{c} \log \frac{h}{c} + b_{\text{height}} 
\ee
Empirically we obtain $a_{\text{height}} \approx -2.5$ from the fit in figure \ref{fig:aparameterfits}.  The $b_{\text{time}}$ and $b_{\text{height}}$ parameters do not fit a simple pattern; we provide some data on these parameters in figure \ref{fig:bparameterdata} in the appendix.  These fits led to the result summarized by equation (\ref{eq:IntroVmaxtmax}) in the introduction, which neglects the small offsets from the $b$-parameters.  We expect that $|\CV_{\text{max}}|$ and $t_{\text{max}}$ are controlled by semiclassical physics (for example, see figure \ref{fig:NonVacSemiVSFull}), so it would be interesting to try to prove these empirical relations using analytic results \cite{Fitzpatrick:2016mjq} on the semiclassical time-dependence.  In principle these results could also be obtained from an AdS calculation involving black holes and deficit angles.

\subsubsection{Probing Exponentially Large Timescales}

Formally, we are interested in high-energy pure states corresponding to BTZ black holes, which have a large entropy $S = \frac{\pi^2}{3} c T_H$ in the large central charge limit where AdS gravity provides a reliable description.  This suggests that timescales of order $e^S$ will be unreachably large.  Nevertheless, by considering either small $c$ or small $T_H$, we can probe order one $S$, and thus reach $t \sim e^S$ within the range of convergence of our numerics.  

\begin{figure}[h]
\begin{centering}
\includegraphics[width=0.98\textwidth]{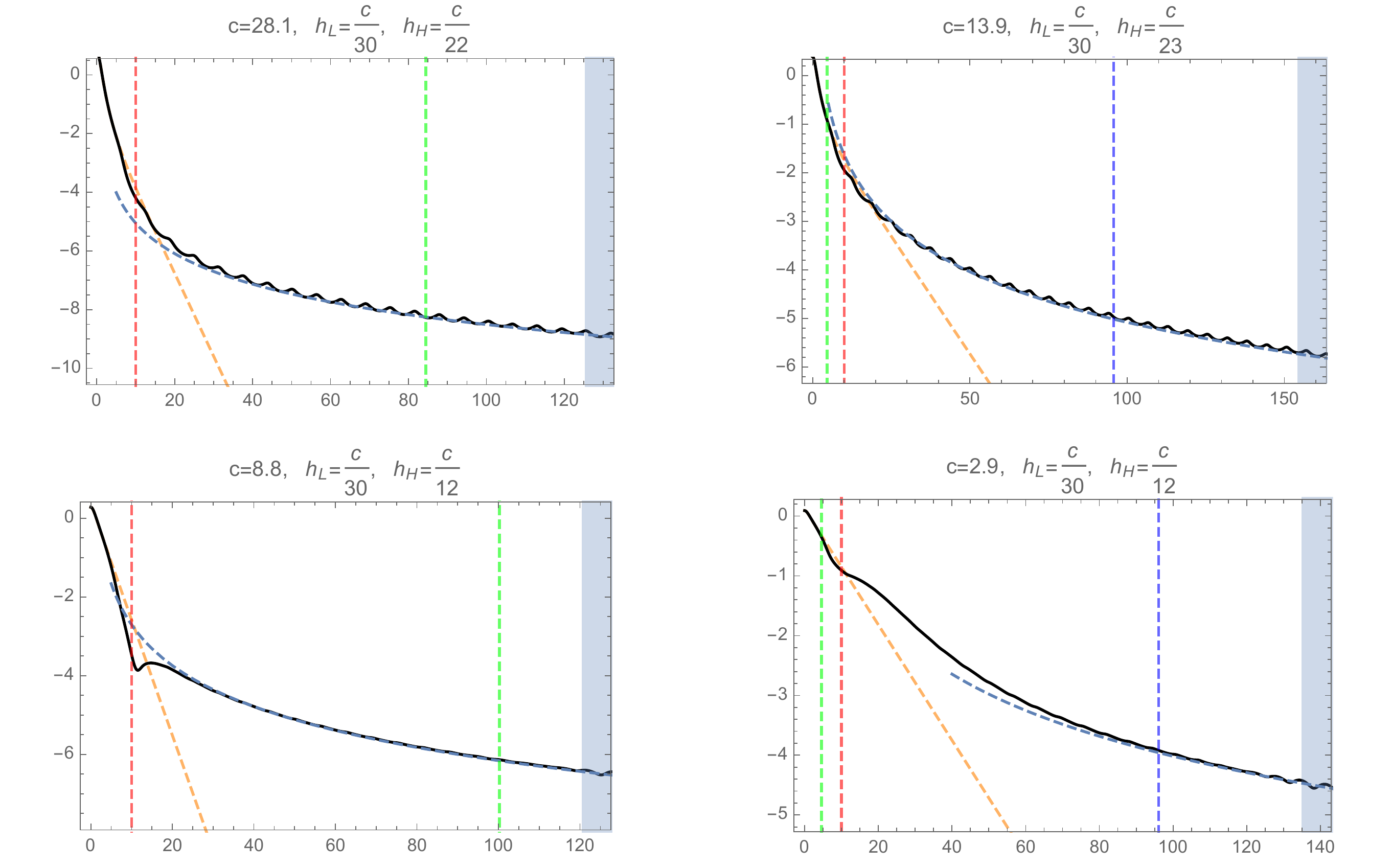}
\caption{ These plots show a variety of parameter choices where the behavior of Virasoro blocks on the timescale $e^S$ (green vertical line), and even $e^{e^S}$ (blue vertical line), are visible.  Yellow lines indicate semiclassical behavior, while the light blue fit corresponds to $t^{-\frac{3}{2}}$.  Recall $S = \frac{\pi^2}{3} c T_H$ with $2 \pi T_H =  \sqrt{\frac{24 h_H}{c} - 1}$, so some plots have relatively large $c$ and small $T_H$, while others have order one $T_H$ but small $c$. In all cases we see that the $t^{-\frac{3}{2}}$ late-time decay persists on these exponentially long timescales.  These plots all display vacuum blocks, but we have found similar behavior with $h > 0$.}
\label{fig:VBlockESEES}
\end{centering}
\end{figure}

In fact, the plot on the bottom-right of figure \ref{fig:VBlock} is already in this regime.  Due to its low temperature of $T_H \approx 0.03$ in AdS units, we have $S \approx 3.3$ so that times of order $e^S \approx 27$ are within the range of convergence.  Thus this plot already suggests that the $t^{-\frac{3}{2}}$ power-law decay persists to exponentially large timescales.  In figure \ref{fig:VBlockESEES} we have displayed four other choices of parameters where timescales of order $e^S$, and even $e^{e^S}$, are visible within the range of convergence.  Two examples have order one $T_H$ and small $c$, while two others have very small $T_H$ and relatively large $c$.  In all cases we see that the $t^{-\frac{3}{2}}$ late-time decay persists on these exponentially large timescales.  This provides good evidence that the heavy-light Virasoro blocks really do decay in this way at very late times.  This means that these blocks are not sensitive to the discreteness of the spectrum in other channels.

\subsection{Power Law Behavior of $q$-Expansion Coefficients}
\label{sec:CoeffPowerLaw}

We have observed an apparently universal late-time power-law behavior in the heavy-light Virasoro blocks $\CV_h(t)$.  One might try to derive this behavior by studying its implications for the $q$-expansion.  In fact, for a large region of parameter space, the $t^{-\frac{3}{2}}$ decay  translates to a power law growth of the coefficients in the $q$ expansion. 

\begin{figure}[H]
\centering{}\includegraphics[scale=0.5]{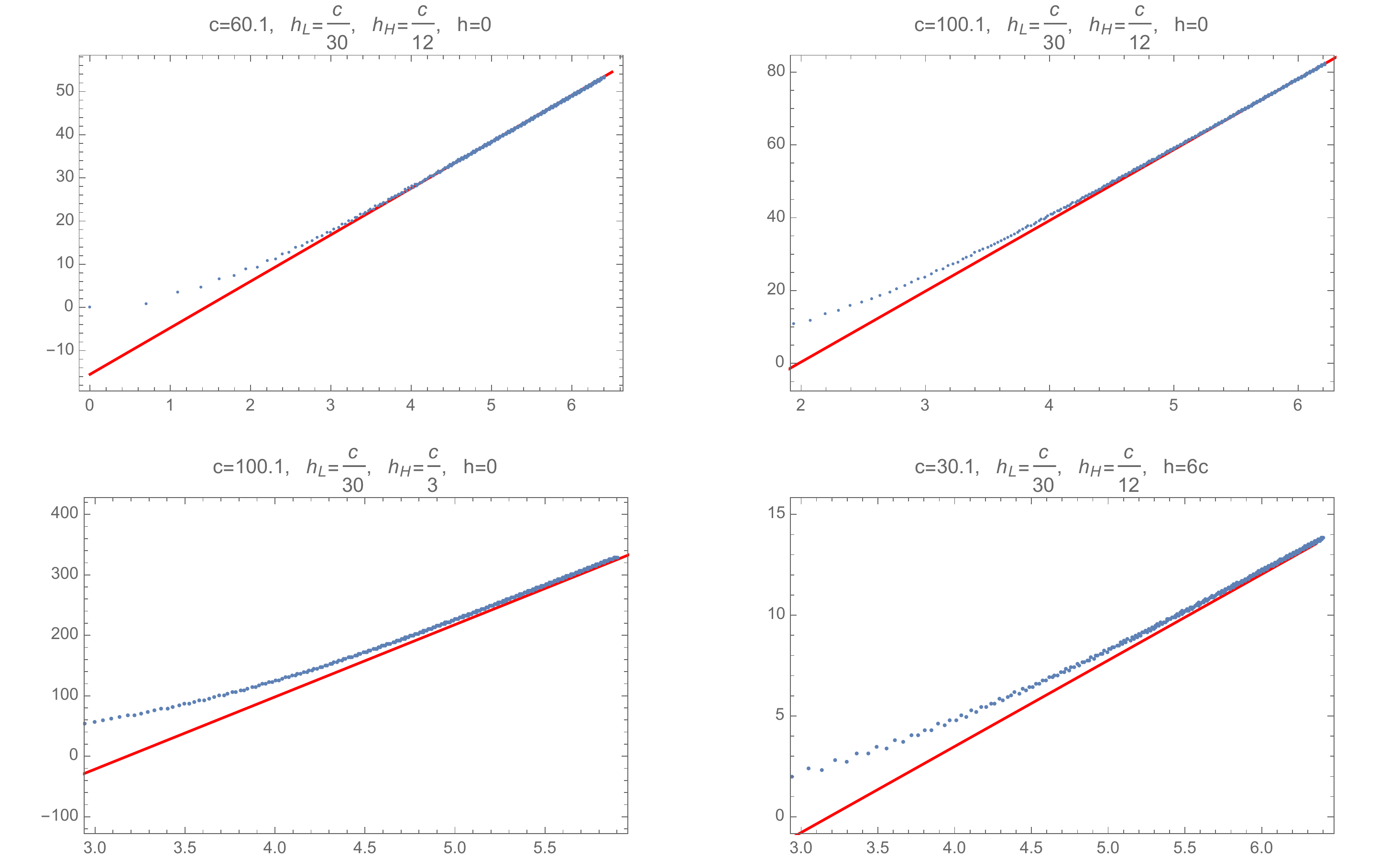}\caption{The behavior of the coefficients of the $q^{2n}$ term in the polynomial $H$ in (\ref{eq:FullVBlock}) compared to the prediction (\ref{eq:CoeffPower}). The horizontal axis is $\log n$ and the vertical axis is $\log c_{n}$, where $c_n$ is the coefficient of $q^{2n}$ in $H$. The red lines are power-laws $a n^s$ with the constant $a$ determined by the fit. \label{fig:Coeffpower}}
\end{figure}

To see this, we note that at late times $q$ approaches $1$ with a rate given by (\ref{eq:LargeTimeqApprox}). This implies that $\theta_{3}(q)\sim\sqrt{t}$, which means that the prefactor in (\ref{eq:FullVBlock}) behaves like  $t^{\frac{1}{2}\left(\frac{c-1}{2}-8(h_{H}+h_{L})\right)}$ at late times. In order to have the entire block decay as $t^{-\frac{3}{2}}$, the polynomial part $H\left(c,h,h_L, h_H ,q\right)$ must  cancel all $c$ and $h_i$ dependence in the prefactor. This means: 
\begin{equation}
H(t)={\displaystyle \sum_{n=0}^{\infty}}c_{n}q(t)^{2n} \ \sim \ t^{4\left((h_{H}+h_{L})-\frac{c}{16}-\frac{5}{16}\right)}\label{eq:PowerSum}
\end{equation}
A power law in the late time behavior of the $H$ can be directly related to the large order behavior of the $q$-expansion coefficients $c_n$.  We find that $c_{n} \sim n^{s}$ with 
\be
\label{eq:CoeffPower}
s=4\left(h_{H}+h_{L} - \frac{c}{16}-\frac{9}{16}\right)
\ee
where $s$ is the dominant power of the coefficient growth, and we are assuming that $H(t)$ does grow at large $t$, which roughly requires $h_H > \frac{c}{16}$. Examples of this behavior are shown in figure \ref{fig:Coeffpower}.  
 If $H(t)$ decays at late times, then there must be cancellations in the sum over $q^n$, and we cannot predict such a simple power-law.

So in addition to directly computing the late time values of the Virasoro blocks, we can test whether the blocks follow the $t^{-\frac{3}{2}}$ decay simply by comparing the coefficients of the $q$-expansion of $H$ to the prediction (\ref{eq:CoeffPower}). This is actually a more efficient method that allows us to access certain regimes, such as larger $c$ and $h_H$ of the parameter space where the direct Virasoro block calculation converges poorly. 

 However, the prediction (\ref{eq:CoeffPower}) is less universal than the $t^{-\frac{3}{2}}$  behavior.  For example, outside the regime where $H(t)$ grows, the coefficients $c_n$ can have alternating signs, so that there are  large cancellations between different terms in the $q$-expansion.  Then the magnitude of the coefficients will no-longer follow the simple pattern depicted in figure \ref{eq:CoeffPower}.  Empirically, another example is when $\frac{h_L}{c}$ is small. In this case the coefficients are pretty small and show complicated irregular behaviors. Examples can be see in figure \ref{fig:CoeffEvolution} in the appendix. Yet in all cases the overall late time behavior of the heavy-light Virasoro blocks is still the $t^{-\frac{3}{2}}$ power law.  

One would hope to derive the power-law behavior $c_n \sim n^s$ using the Zamolodchikov recursion relations.  Unfortunately, it appears that this behavior arises from a large number of cancellations between much larger terms.  Thus we leave this problem to future work.

\subsection{Implications for Information Loss and the Bootstrap}

In the semiclassical limit, heavy-light Virasoro blocks decay exponentially at late times.  We do not expect that perturbative corrections in $G_N = \frac{3}{2c}$ will alter this conclusion, and to first order this has been demonstrated explicitly \cite{Fitzpatrick:2015dlt}.  Thus the late time power-law behavior of the exact blocks represents a non-perturbative correction that ameliorates information loss (insofar as information loss is tantamount to late-time decay).  However, since the Virasoro blocks continue to decay, albeit much more slowly, this effect does not solve the information loss problem.  For this we need an infinite sum over Virasoro blocks in the $\CO_L \CO_L$ OPE channel.\footnote{Of course we are assuming that we are dealing with a chaotic large $c$ theory, rather than e.g. a rational CFT.  For special values of the external dimensions and $c$, such as those corresponding to degenerate external operators, the individual Virasoro blocks may not decay at late times.}  

Let us examine the correlator as a sum over blocks from the point of view of the bootstrap equation \cite{FerraraOriginalBootstrap1, PolyakovOriginalBootstrap2, Rattazzi:2008pe}.  This equation dictates that\footnote{We are being schematic to emphasize the time dependence.  One should define $z = 1 - e^{-t + i \phi}$ and $\bar z = 1 - e^{-t - i \phi}$ in the Euclidean region, and then analytically continue $t \to it$, so that both channels depend on the coordinates $t$ and $\phi$ pictured in figure \ref{fig:TimelikeSeparation}.  We are suppressing these details.}
\be \label{eq:EnergyDecomp}
\< \CO_H(\infty) \CO_L(t) \CO_L(0) \CO_H(-\infty) \> = \sum_E   \lambda^2_{LH}(E) e^{i E t}
= \sum_{h, \bar h} P_{h, \bar h} \CV_h(t) \bar \CV_{\bar h}(t)
\ee
Here we have equated a sum over energies in the $\CO_H \CO_L$ OPE channel with a sum over heavy-light Virasoro blocks in the $\CO_L \CO_L$ OPE channel.  In $d > 2$ dimensions this equation would be meaningless at large $t$, because we would be well outside the regime of convergence of the OPE expansion on the right-hand side.  Remarkably, as discussed in section \ref{sec:Kinematicsandq}, the Virasoro block decomposition converges for all values of $t$, so it is possible to try to `solve' for the coefficients $P_{h, \bar h}$ by equating the large $t$ behavior of both sides.  More generally, one could take the limit $|q| \to 1$ with various phases for $q$ and derive new, potentially analytic regimes for the bootstrap (this is non-trivial because it could enable a partial analytic treatment without requiring a complete solution to the bootstrap equation).  The only obvious obstruction to this procedure is that we do not have simple analytic formulas for the Virasoro blocks in such limits.  

As we have already noted, equation (\ref{eq:EnergyDecomp}) can only be satisfied at late times if we have an infinite number of Virasoro blocks contributing on the right-hand side.  Such infinite sums are compulsary in order to reproduce conventional OPE limits \cite{Cardy:1986ie, Pappadopulo:2012jk, Fitzpatrick:2012yx, KomargodskiZhiboedov}.  But it is easy to see that the Cardy formula and the asymptotic expectations on $P_{h, \bar h}$ from Euclidean crossing or the light-cone OPE limit are insufficient to account for the late-time behavior.  The reason is that conventional arguments require the large $h, \bar h$ terms in equation (\ref{eq:EnergyDecomp}) to reproduce either the identity (vacuum) or perhaps the contribution of low dimension or low twist operators in the crossed channel. These would correspond to the very small $E$ region of $\lambda_{LH}(E)$.  But the late time behavior arises from the collective contributions of $\sim e^S$ states with large $E \sim h_H + \bar h_H$, not from the small $E$ states.\footnote{Here we are imagining subtracting off the contributions from the expectation values $\<\CO_H | \CO_L | \CO_H \>$. These are generically expected to be exponentially suppressed \cite{Kraus:2016nwo}  in holographic CFT$_2$.}

In this regard there is an amusing connection with Maldacena's original discussion \cite{Maldacena:2001kr} of the large time behavior.  He suggested that in a black hole background, contributions from the vacuum, corresponding to the $E=0$ term in equation (\ref{eq:EnergyDecomp}), might resolve the information loss problem.  But the vacuum in the $\CO_H \CO_L$ OPE channel just corresponds with the Cardy-type growth (or more precisely OPE convergence \cite{Pappadopulo:2012jk} type growth) of $P_{h, \bar h}$.  So this simple OPE convergence growth fails to account for the late time behavior for the same reason that Maldacena's suggestion did not resolve the information loss problem.
 
In summary, the late-time bootstrap equation (\ref{eq:EnergyDecomp}) cannot be solved without providing a more refined asymptotic formula for $P_{h, \bar h}$ at large $h, \bar h$.  However, it does not appear that a discrete spectrum in the $\CO_L \CO_L$ channel is required to obtain the correct late-time behavior.  We will not pursue this in detail since we only have some rough empirical information about the behavior of $\CV_h(t)$, but it might be interesting to study this bootstrap equation for the case of the partition function \cite{Dyer:2016pou} where the Virasoro characters are known in closed form.

\section{Euclidean Breakdown of the Semiclassical Approximation}
\label{sec:ForbiddenSingularitiesandReconstruction}

\subsection{Some Philosophy}

Eventually, we hope to learn about bulk reconstruction -- and its limitations -- by comparing exact CFT correlators to their semiclassical approximations.  It is not clear whether this is possible, even in principle, due to ambiguities in the reconstruction process associated with bulk gauge redundancies (see e.g. \cite{Jafferis:2017tiu} for a recent discussion).  For now we will take a very instrumental approach, or in other words, we will try to `shut up and calculate' some potentially interesting observables.  

The information paradox pits local bulk effective field theory in the vicinity of a horizon against quantum mechanical unitarity.  But in the strict semiclassical limit, information is lost and the (approximate) CFT correlators agree precisely with perturbative  AdS field theory or string theory.  Thus one would expect that bulk reconstruction should be possible in this approximation, since we have allowed the local bulk theory to `win' the fight, at the expense of unitarity.\footnote{This suggests that solving the reconstruction problem in the strict semiclassical limit should not have much to do with the information paradox or the existence of firewalls \cite{Almheiri:2012rt}, except insofar as it is a first step towards the problem of bulk reconstruction from the data and observables of the exact CFT.  As an alternative perspective, one might claim that even in the semiclassical limit reconstructing black hole interiors is impossible because firewalls are completely generic.}  

But even in the semiclassical limit, bulk reconstruction has been controversial \cite{Hamilton:2005ju, Bousso:2012mh, Leichenauer:2013kaa, Morrison:2014jha}.  On an intuitive level, this is because correlators at infinity must have exponential sensitivity to `observe' physics near or behind a black hole horizon.  At an instrumental level, this means that there may be obstructions to the existence of smearing functions mapping boundary to bulk observables.  These issues can be avoided by going to momentum space \cite{Papadodimas:2012aq, Papadodimas:2013jku}, or perhaps via an appropriate analytic continuation \cite{Hamilton:2005ju} or cutoff procedure \cite{Morrison:2014jha}.

Another elementary issue with semiclassical bulk reconstruction is  pictured in figure \ref{fig:PenroseIngoingOutgoing}.  The problem is that only the ingoing modes behind the horizon can be reconstructed in an obvious way from the degrees of freedom of a single CFT \cite{Hamilton:2005ju}.  This can be understood by considering the extended AdS-Schwarzschild spacetime, or simply by studying  Rindler space.  Field theory degrees of freedom behind the horizon appear as a linear combination of modes from the left and right `wedges', but in a single-sided black hole, only one asymptotic region is present.

If the goal is simply to compute correlators behind the horizon of a single-sided black hole, then there is a naive, instrumental way to obtain outgoing modes.  One can obtain correlators that behave like those of the other asymptotic region by analytically continuing \cite{Hamilton:2005ju} CFT operators $\CO(t, x)$ in Euclidean time to $\tilde \CO(t, x) \equiv \CO \left(t + \frac{i \beta}{2}, x \right)$.  This procedure has an important flaw -- operators on opposite sides of the black hole should  commute, but $\CO$ and $\tilde \CO$ may not. Nevertheless, we can force $\CO$ and $\tilde \CO$ to commute (by definition) if we choose an appropriate but ad hoc analytic continuation procedure for correlators involving $\CO$ and $\tilde \CO$.  Conceptually, this does not seem to be an improvement on state-dependent mirror operators \cite{Papadodimas:2012aq, Papadodimas:2013jku}, which represent a modification of quantum mechanics.  In fact, our procedure implements its own form of state-dependence, since the analytic continuations will depend on all of the other local operators inserted into the correlator.  However, the prescription does have the simple advantage of being relatively precise and unambiguous.

In any case, we are led to a very simple question -- do the correlators of operators like $\CO \left(t + \frac{i \beta}{2}, x \right)$ receive large non-perturbative corrections?  Do the semiclassical Virasoro blocks provide a good approximation to the exact blocks with these kinematics?  

\subsection{Forbidden Singularities and Thermofield Doubles}

The questions raised in the previous section can be explored using the methods of this paper.  They are also closely related to observations about information loss \cite{Fitzpatrick:2016ive}.  Finite-temperature correlation functions must satisfy the KMS condition, which for identical operators just means that $\< \CO(t, x) \CO(0) \>_\beta$  must be periodic in Euclidean time with period $\beta$.  It has been shown that in the large central charge limit with $h_H > \frac{c}{24}$, heavy-light Virasoro blocks appear thermal.\footnote{The vacuum block is exactly periodic.  The general case in equation (\ref{eq:HeavyLightBlocks}) would be periodic except for the branch cuts of the hypergeometric function, but these do not obstruct the KMS condition for the full correlator, and are compatible with the Virasoro block decomposition of correlators obtained from BTZ black hole backgrounds \cite{Fitzpatrick:2015zha}.}  Since the 4-point correlator has an OPE singularity
\be
\< \CO_H(0) \CO_L(z) \CO_L(1) \CO_H(\infty) \> = \frac{1}{(1-z)^{2h_L}} + \cdots
\ee
as $z \to 1$, in the heavy-light semiclassical limit, it will also have singularities at $z_n = 1 - e^{n \beta}$ for all integers $n$.   

\begin{figure}[t]
\centering{}\includegraphics[]{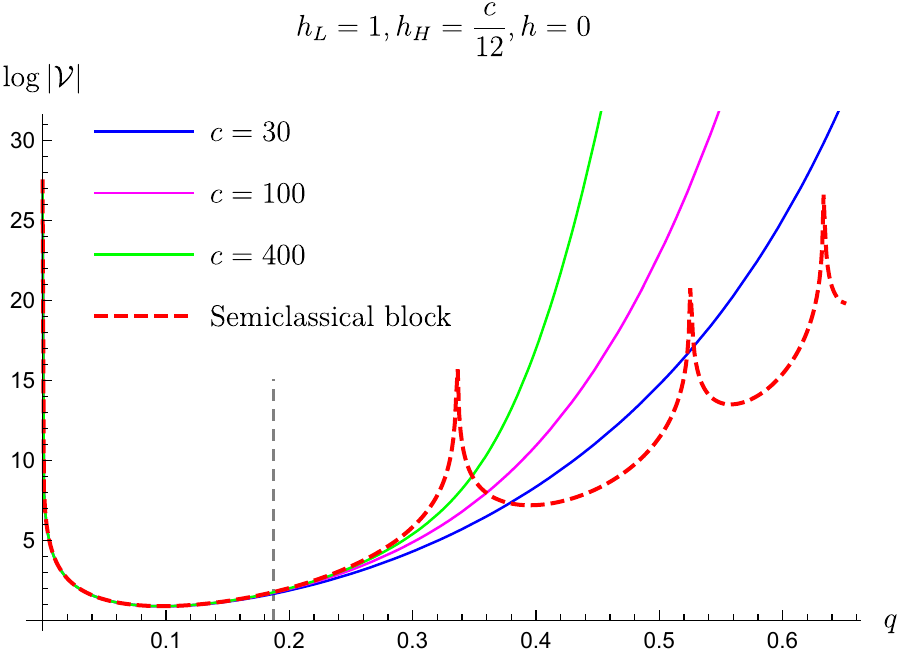}
\caption{In this plot, we compare the exact and semiclassical blocks. One can see that at the positions of the semiclassical forbidden singularities, the exact blocks are smooth. Fixing $h_L$ and $\frac{h_H}{c}$ as we increase $c$, the exact blocks approach the semiclassical block in the region between the origin and the first forbidden singularity.  However, beyond the first forbidden singularity the exact blocks deviate greatly as we increase $c$.  This indicates that we have passed a Stokes line (emanating from the forbidden singularity) and some other semiclassical saddle dominates the exact blocks in the large $c$ limit.  The gray line is the position of $t =  \frac{i \beta}{2}$.}
\label{fig:forbidden}
\end{figure}

While such singularities are permissible for correlators in the canonical ensemble, they are forbidden \cite{Maldacena:2015iua, Fitzpatrick:2016ive} from 4-point correlators of local operators in unitary CFTs.  They are also forbidden from individual Virasoro blocks at finite central charge \cite{Fitzpatrick:2016ive, Fitzpatrick:2016mjq}.  Thus exact Virasoro blocks completely disagree with their semiclassical counterparts at $z_n = 1 - e^{n \beta}$, the locations of the singularities.  So to summarize, we know that the exact and semiclassical blocks match at $z = 0$, and completely disagree at $z = 1 - e^{n \beta}$ for $n \neq 0$.  Thus it is natural to wonder whether the semiclassical blocks are a good approximation at $z = 1 - e^{-\frac{\beta}{2} -it}$, which corresponds to the location of $\CO(t + \frac{i \beta}{2})$.  More generally we would like to understand the kinematical regimes where the (leading) semiclassical approximation breaks down.

\begin{figure}[t]
\hfill\includegraphics[]{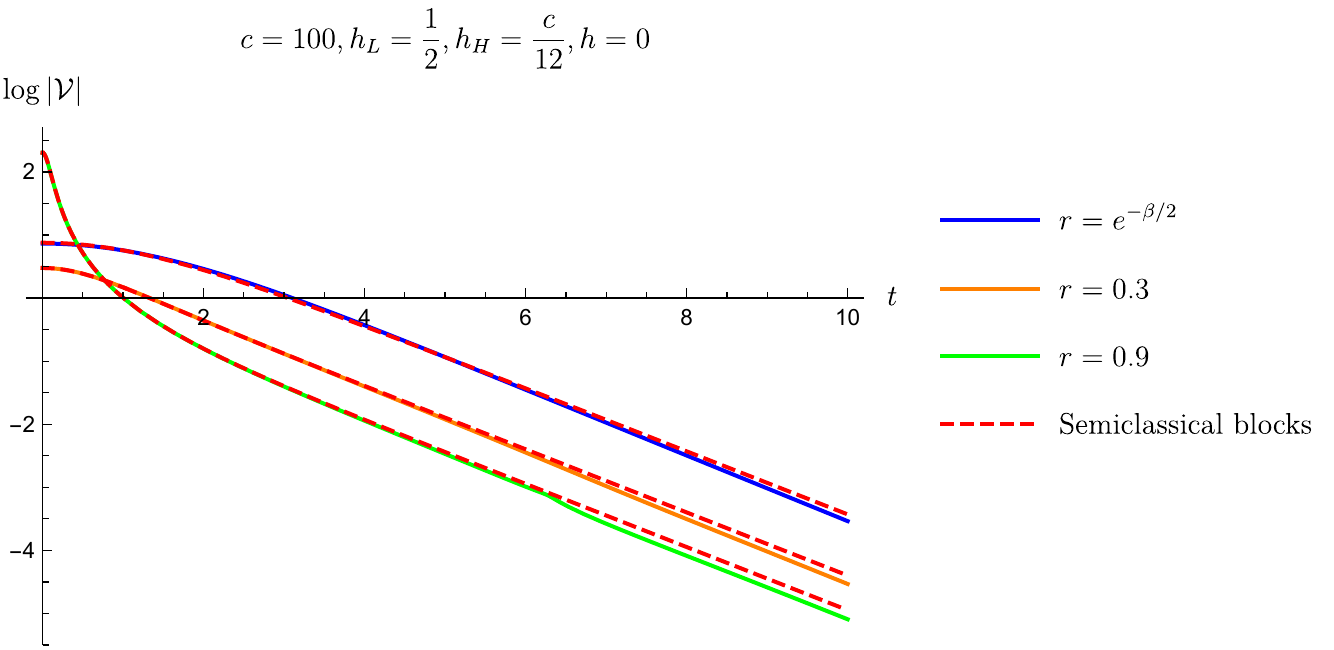}
\caption{In this figure we compare the semiclassical and exact blocks associated with $\CO(t)$ and $\CO(t + \frac{i \beta}{2})$.  The plot suggests that the semiclassical approximation remains valid for correlators of $\CO(t + \frac{i \beta}{2})$.  We implement time dependence via $z = 1 - r e^{-it}$ and so a shift by $\frac{i \beta}{2}$ simply corresponds to a different choice of $r$.  Corresponding trajectories in the unit $q$ disk are pictured in figure \ref{fig:qoft}.   Apparently the semiclassical approximation works well at $t + \frac{i \beta}{2}$.   }
\label{fig:betaover2}
\end{figure}

We observe from figure \ref{fig:forbidden} that as expected, the exact Virasoro blocks do not have forbidden singularities.  Nevertheless one might have expected to see bumps or local maxima at $z_n = 1 - e^{n \beta}$, whereas the exact correlator simply grows as a function of $z \in [0,1)$.  In fact local maxima are prohibited because the exact blocks are analytic functions of $q$ and $z$ away from the true OPE singularities.\footnote{Moreover it is not too surprising that a finite series expansion of the exact blocks simply grows in the region where the semiclassical blocks have forbidden singularities.  For example, the finite-order series expansion of a function like $\frac{1}{(1-x)^2(2-x)^2}$ will grow monotonically on the positive real $x$-axis; one can only see the correct behavior on $x \in (1,2)$ by summing the full series and analytically continuing around $x=1$.}  Thus the semiclassical approximation breaks down badly beyond the first forbidden singularity. 

We compare the exact and semiclassical blocks at finite time in figure \ref{fig:betaover2}.   We see that the semiclassical blocks remain a good approximation to correlators of $\CO(t + \frac{i \beta}{2})$ as long as we avoid the long-time region of $t \propto S$ that was discussed in section \ref{sec:LateTime}.  In particular,  there is not a significant difference between the quality of the semiclassical approximation to correlators of $\CO(t + \frac{i \beta}{2})$ and $\CO(t)$.  The most naive interpretation of this fact is that non-perturbative quantum gravitational effects do not obstruct local physics across the horizon of pure, energy-eigenstate black holes.  A qualitatively similar conclusion was reached for late-time deviations \cite{Kabat:2014kfa} from the semiclassical limit.  This result was also anticipated by the analytic analysis of \cite{Fitzpatrick:2016ive}, which only suggested large non-perturbative corrections within $\frac{1}{\sqrt{c}}$ of the forbidden singularites.  In the next section we will discuss that analysis and compare it with our numerical results.

\subsection{Fate of the Semiclassical Approximation from Analytics and Numerics}
\label{sec:FateSemiclassical}

\begin{figure}[t]
\centering{}\includegraphics[width=0.9\textwidth]{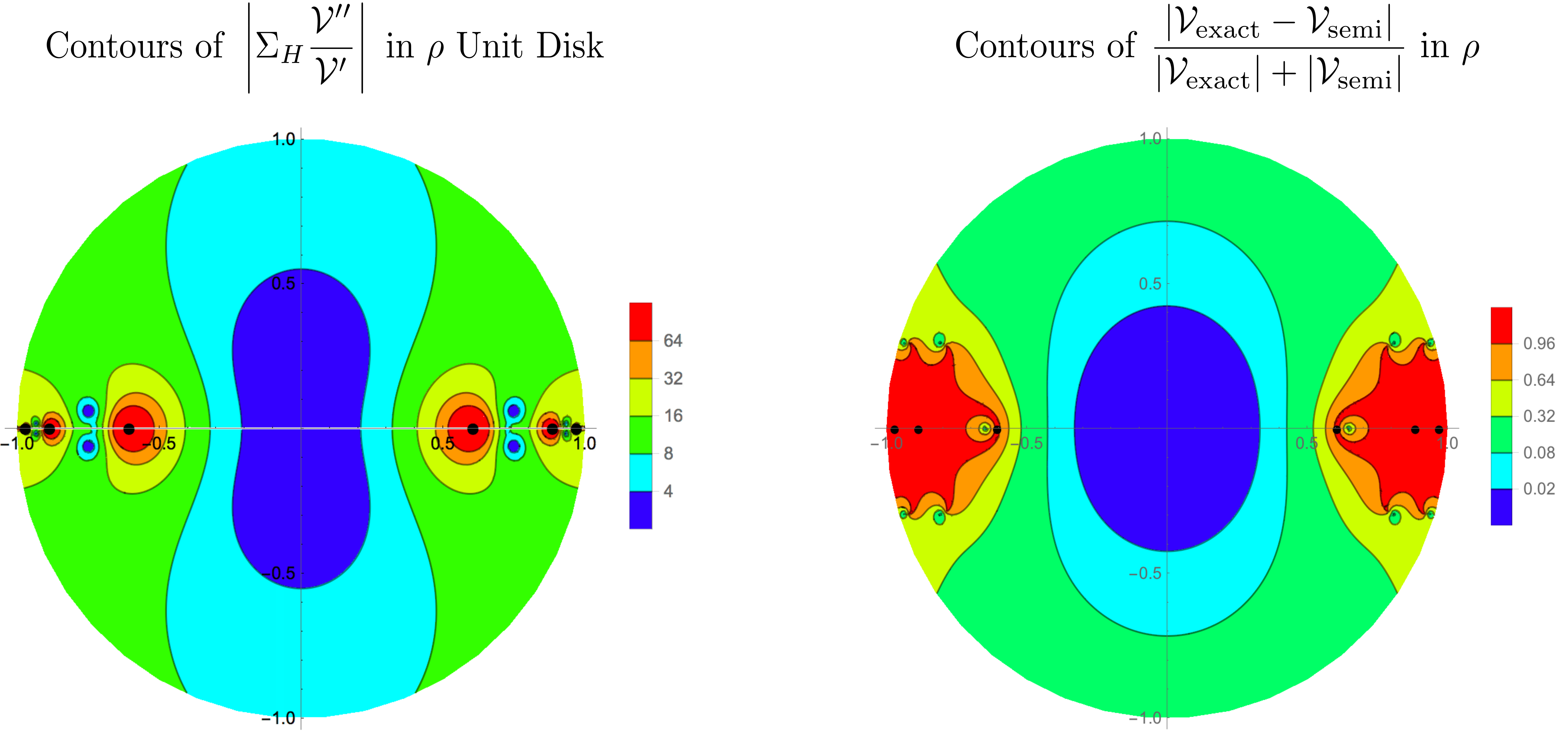}
\caption{The figure on the left shows a contour plot of the function $|\Sigma_{H} \frac{\CV''}{\CV'}|$ from equation (\ref{eq:2ndOrderVacDiffEq}) in the $\rho$ unit disk with $h_L = 1$ and $h_H = \frac{c}{4}$.  The figure on the right is the deviation of the exact and semiclassical Virasoro vacuum blocks with the same parameters and $c = 60$.  The positions of the forbidden singularities are indicated with black dots. The plot on the left can be viewed as a kind of analytic prediction for the deviation plotted on the right. }
\label{fig:ContoursSigmaFunction}
\end{figure}

We do not have to rely entirely on numerics to explore the regime of validity of the semiclassical limit.  It has been shown that the vacuum block's forbidden singularities have a universal resolution due to non-perturbative effects in central charge.  Specifically, the heavy-light  vacuum block (with $h_L$ and $\frac{h_H}{c}$ held  fixed at large $c$) should obey an approximate differential equation \cite{Fitzpatrick:2016ive}
\be
\label{eq:2ndOrderVacDiffEq}
h_L g_H(\tau) \frac{\CV(\tau)}{\CV'(\tau)} - 1\  &=&   \ \frac{6}{c} \Sigma_{H}(\tau) \frac{\CV''(\tau)}{\CV'(\tau)} 
\ee
where $\tau = -\log(1-z)$ is a Euclidean time variable, and this equation neglects terms of order $1/c^2$ and higher as well as effects that are less singular near the forbidden singularities.  We provide the functions $g_H$ and $\Sigma_H$ in appendix \ref{app:DiffEqforVac}.  This differential equation also predicts  \cite{Fitzpatrick:2016ive} that the semiclassical vacuum block will receive large non-perturbative corrections after a Lorentzian time of order $\frac{S_{BH}}{h_L T_H}  \propto \frac{c}{h_L}$. That prediction was corroborated  in section \ref{sec:LateTime}.

Neglecting the term proportional to $\frac{1}{c}$  on the right-hand side, equation (\ref{eq:2ndOrderVacDiffEq}) is solved by the semiclassical heavy-light vacuum block.  But when the right-hand side of this equation becomes large, non-perturbative effects come into play, resolving the forbidden singularities. We plot contours of the function $|\Sigma_{H} \frac{\CV''}{\CV'}|$ for $h_L = 1$ in figure \ref{fig:ContoursSigmaFunction}.  We see that this function becomes large and makes important contributions in the immediate vicinity of the forbidden singularities, though at sufficiently large $c$ the right-hand side of equation (\ref{eq:2ndOrderVacDiffEq}) will remain small a finite distance away from these singularities.  At a more detailed level, the function $|\Sigma_{H} \frac{\CV''}{\CV'}|$  can be compared directly to the deviation of the numerical and semiclassical vacuum block. We plot contours of the ratio of the exact and semiclassical blocks in the $\rho$ unit disk, corresponding to the entire Euclidean $z$-plane in figure \ref{fig:ContoursSigmaFunction} (recall that we compared various kinematic variables in figures \ref{fig:zrhoqBranchCuts} and \ref{fig:Constantq}).

Our numerical results demonstrate that the semiclassical approximation breaks down  in a finite region enclosing the forbidden singularities.   We believe this phenomenon occurs because Stokes and anti-Stokes lines (for review see e.g. \cite{Witten:2010cx}) emanate from the forbidden singularities, as has been demonstrated for the correlators of degenerate operators \cite{Fitzpatrick:2016ive}.  As we cross Stokes lines, the coefficients of semiclassical saddles change by discrete jumps.  Across anti-Stokes lines saddles exchange dominance.  

\begin{figure}
\centering{}\includegraphics[width=0.9\textwidth]{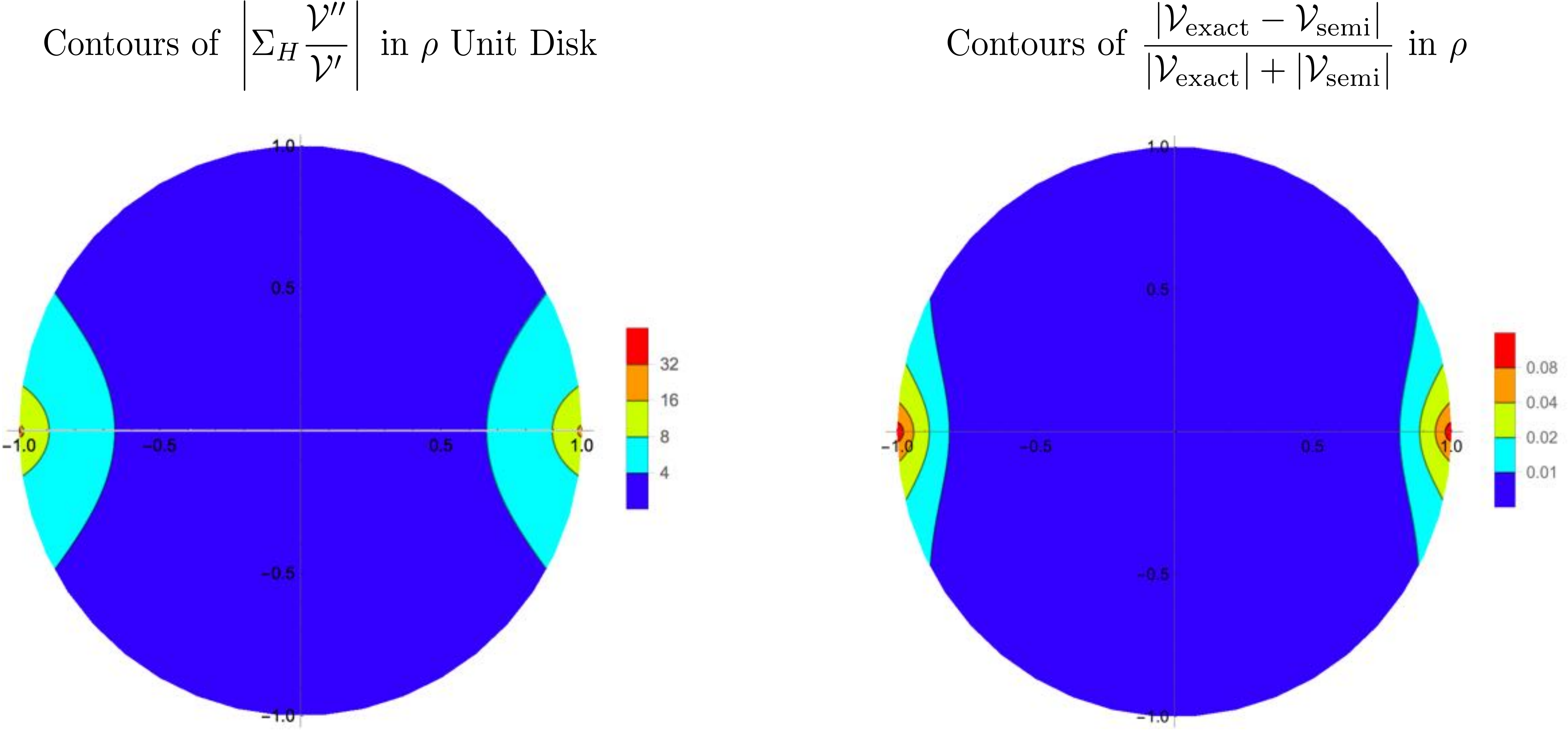}
\caption{The figure on the left shows a contour plot of the function $|\Sigma_{H} \frac{\CV''}{\CV'}|$ from equation (\ref{eq:2ndOrderVacDiffEq}) in the $\rho$ unit disk with $h_L = 1$ and $h_H = \frac{c}{30}$.  In this case $h_H < \frac{c}{24}$, so the heavy-light block does not include a black hole -- instead it corresponds to a light probe interacting with a deficit angle in AdS$_3$.  Thus there are no forbidden singularities, and the semiclasssical approximation is reliable in a much larger region as compared to figure \ref{fig:ContoursSigmaFunction} (note the difference in scales).  The figure on the right is the deviation of the exact and semiclassical Virasoro vacuum blocks with the same parameters and $c = 60$.  The plot on the left can be viewed as a kind of analytic prediction for the deviation plotted on the right. }
\label{fig:NoBHContours}
\end{figure}

Near the OPE configuration $z \propto \rho \propto q \approx 0$ where the light operators collide, a special `original' semiclassical saddle dominates the large $c$ limit \cite{Fitzpatrick:2016ive} of the Virasoro blocks.  But in a finite region near the forbidden singularities, different semiclassical saddles \cite{Fitzpatrick:2016mjq} can come to dominate, and the original saddle may become sub-leading.  In other words, analytic continuation in the kinematic variables does not commute with the large $c$ limit.  Non-perturbative effects can dramatically alter the behavior of CFT$_2$ correlation functions with these kinematics, supplanting the naive semiclassical limit and the perturbation expansion around it.  

It would be fascinating if the black hole interior depends in some way on the behavior of CFT correlation functions in these regimes.  Note that when $h_H < \frac{c}{24}$, so that the heavy background state does not correspond to a black hole, the original semiclassical approximation remains good throughout the Euclidean region. We demonstrate this explicitly in figure \ref{fig:NoBHContours}.  So the breakdown of the semiclassical limit exhibited in figure \ref{fig:ContoursSigmaFunction} really does depend on the presence of a black hole, and is not a general feature of all Virasoro blocks at large central charge.

\section{Discussion}

We would eventually like to resolve the black hole information paradox by doing the right calculation.  In the context of AdS/CFT, this means discerning under what circumstances, if any, bulk reconstruction is possible near and behind black hole horizons.  

If firewalls \cite{Almheiri:2012rt} are completely generic, or if bulk reconstruction is sufficiently ambiguous, then this could be a fools errand.    But even in this case, one can still hope for a more constructive argument rather than various reductio ad absurdums \cite{Almheiri:2013hfa}.  For example, one would like to reconstruct the `experience' of a collapsing spherical shell, and explicitly compute the timescale beyond which subsequent infallers will not see a smooth (or well-defined) geometry.

But let us imagine that the strict semiclassical limit is not misleading and black holes often have smooth interiors.  In this case, violations of bulk locality should arise from the difference between computations in the semiclassical limit and the exact CFT observables (or perhaps meta-observables).  This sort of approach has already been successfully pursued in the context of local bulk scattering \cite{Maldacena:2015iua}.  We have identified gross differences between exact and semiclassical CFT correlators in both the late Lorentzian time and the Euclidean regime.    These do not seem to affect a certain naive bulk reconstruction algorithm, but perhaps they do afflict more sophisticated methods yet to be developed.   Hopefully we have done some of the right calculations but do not yet know how to give them the right interpretation.  In the case of quantum mechanics and QFT, we were in that sort of boat for decades.

\section*{Acknowledgments}
 
We would like to thank Ibou Bah, Ethan Dyer, Tom Faulkner, A. Liam Fitzpatrick, Guy Gur-Ari, Daniel Harlow, Diego Hofman, Tom Hartman, David Kaplan, Alex Maloney, Sam McCandlish, Suvrat Raju, Douglas Stanford, Junpu Wang, and the participants of the JHU Conference on the Bootstrap and Quantum Gravity for discussions. We also thank A.L. Fitzpatrick for comments on the draft.  JK, HC, and CH  have been supported in part by NSF grant PHY-1454083.  We were also supported in part by the Simons Collaboration Grant on the Non-Perturbative Bootstrap.  

\appendix
\section{Details of Recursion Relations and Our Algorithm}
\label{app:ZRecursion}
In this appendix we will present more details about Zamolodchikov's recursion relations and the algorithm we used to compute with them.

\subsection{Zamolodchikov's Recursion Relations}

There are actually two Zamolodchikov recursion relations, based on viewing the Virasoro blocks as either a sum over poles in the central charge $c$ or the intermediate state dimension $h$.  The latter is more powerful and will be our focus. 

The Virasoro block of the four-point function $\left<\CO_1(0)\CO_2(z)\CO_3(1)\CO_4(\infty)\right>$ with central charge $c$, external dimensions $h_i$ and intermediate dimension $h$ takes the following form
\begin{equation}
\mathcal{V}_{h,h_{i},c}(z)=\left(16q\right)^{h-\frac{c-1}{24}}z^{\frac{c-1}{24}-h_{1}-h_{2}}\left(1-z\right)^{\frac{c-1}{24}-h_{2}-h_{3}}[\theta_{3}\left(q\right)]^{\frac{c-1}{2}-4\sum_{i=1}^{4}h_{i}}H\left(c,h_{i},h,q\right),
\end{equation} \label{eq:blockV}
where 
\begin{equation}
	q=e^{i\pi \tau}, \ \ \tau={i \frac{K(1-z)}{K(z)}},
\end{equation}
and the inverse transformations is 
\begin{equation}
	z=\left(\frac{\theta_2(q)}{\theta_3(q)}\right)^4.
\end{equation}
 If we parametrize the central charge $c$, the external operator dimensions $h_i$ and the degenerate operator dimensions $h_{mn}$ as follows
\begin{equation}
	c=13+6\left(b^2+\frac{1}{b^2}\right),\quad  h_{i}=\frac{1}{4}\left(b+\frac{1}{b}\right)^2-\lambda_i^2,\quad h_{m,n}=\frac{1}{4}\left(b+\frac{1}{b}\right)^2-\lambda_{m,n}^2,
	\end{equation}
with 
\begin{equation}
\lambda_{m,n}=\frac{1}{2}\left(\frac{m}{b}+nb\right),
\end{equation}
then the function $H\left(b,h_i,h,q\right)$ can be calculated using the following recursion realtion
\be \label{eq:recursionH}
H(b,h_i, h, q) = 1 + \sum_{m,n\ge1} \frac{q^{mn} R_{m,n}}{h - h_{m,n}} H(b,h_i, h_{m,n} + mn, q),
\ee
where $R_{m,n}$ is given by 
\begin{equation}\label{eq:Rmn}
R_{m,n}=2\frac{\prod_{p,q}\left(\lambda_1+\lambda_2-\lambda_{p,q}\right)\left(\lambda_1-\lambda_2-\lambda_{p,q}\right)\left(\lambda_3+\lambda_4-\lambda_{p,q}\right)\left(\lambda_3-\lambda_4-\lambda_{p,q}\right)}{\prod_{k,l}'\lambda_{k,l}},
\end{equation}
and the ranges of $p,q,k,$ and $l$ are:
\begin{align*}
p&=-m+1,-m+3,\cdots,m-3,m-1,\\
q&=-n+1,-n+3,\cdots,n-3,n-1,\\
k&=-m+1,-m+2,\cdots,m,\\
l&=-n+1,-n+2,\cdots,n. 
\end{align*}
The prime on the product in the denominator means
that $\left(k,l\right)=\left(0,0\right)$ and $\left(k,l\right)=(m,n)$ are excluded.
Note that our definition of $\lambda_{p,q}$ differs by a factor of $-\frac{i}{2}$
from the original paper.

In each iteration of the recursion relation \ref{eq:recursionH}, the only thing that changes is the value of the intermediate state dimension $h\rightarrow h_{m,n}+mn$, which only depends on the values of $m$ and $n$. For simplicity we'll omit the arguments and denote $H(b,h_i,h,q)$ as $H$ and $H(b,h_{m,n}+mn,h_i,q)$ as $H_{m,n}$ in the following discussion.

This recursion relation was derived by viewing the Virasoro block $\CV_h$ as a function of the intermediate dimension $h$, so it can be written as a remainder term that survives when $h\rightarrow \infty$ plus a sum over poles at $h=h_{m,n}$, where $h_{m,n}$ are the dimensions of the degenerate operators. The prefactor in front of $H$ in \ref{eq:blockV} is the $h\rightarrow \infty$ limit of $\CV_h$, as can be derived from \cite{ZamolodchikovRecursion, Zamolodchikovq, Zamolodchikov:1986gh}. The reason that $\CV_h$ has poles at $h=h_{m,n}$ is because of the existence of the null-operator (whose norm is zero) at level $mn$ of the descendants of $\CO_{h_{m,n}}$, which usually will make $\CV_h$ diverge when $h\rightarrow h_{m,n}$.\footnote{This is easy to see by writing $\CV_h$ as a sum over contributions from the states in the Verma Module of $\CO_h$. In this sum, we need to orthogonalize the states, but the zero norm of the null-state will appear as a denominator in this process, which causes the divergence.} The residue of the pole at $h_{m,n}$ will be proportional to the block $\CV_{h_{m,n}+mn}$ with  intermediate operator being the null-operator with dimension $h_{m,n}+mn$. Thus, these residues will have high powers of $q$, which accounts for the $q^{mn}$ factor in front of $H_{m,n}$ and naturally makes the Virasoro block $\CV_h$ a series expansion in $q$.  
   
The numerator of the factor $R_{m,n}$ is constructed such that it vanishes when $\CO_{1}$ (or $\CO_{3}$) belongs to the set of operators allowed by the fusion rule of $\CO_2\CO_{h_{m,n}}$ (or $\CO_4\CO_{h_{m,n}}$). The denominator of $R_{m,n}$ comes from the norm of the null-state when $h\rightarrow h_{m,n}$ (factoring out $h-h_{m,n}$); as far as we know, although it has passed numerous checks, it's never been derived from first principles. 
 
\subsection{Algorithm}
In this paper, we only consider the case that $h_1 = h_2 =h_L$ and $h_3 = h_4 = h_H$. Under this circumstance, $R_{m,n}$ becomes directly proportional to $\lambda_{p,q}^2$, so $R_{m,n} = 0$ whenever $(m,n)$ are both odd, because $(p,q)$ can then be $(0,0)$. This means that every $H_{m,n}$ with odd $mn$ is also zero, as every term contributing to it contains at least one $R_{m_l,n_l}$ with odd $m_ln_l$. As a consequence of this, only even powers of $q$ ever appear, and there's no need to compute anything with odd $mn$. This provides some simplification for the calculation, but it's easy to generalize the following discussion to the case that all $h_i$s are different.

Now we turn to the algorithm we used to compute the recursion relation. The main idea is to sort every contribution to the functions $H$ and $H_{m,n}$ by its order in $q$. By doing this from the beginning of the computation, we are able to use lower-level terms as partial sums for the higher-level terms, saving a great deal of computation.

Denote the coefficient of $q^k$ in any function $f$ as $f^{(k)}$. Then the recursion relation for the coefficients of $q^k$ in the function $H$ is
\begin{equation}\label{eq:recursion1}
	H^{(k)} = \sum_{i=2}^k \sum_{\substack{l=1\\m_ln_l=i}}^{\text{div}(i)} \frac{R_{m_l,n_l}}{h - h_{m_l,n_l}} H_{m_l,n_l}^{(k-i)},
\end{equation}
where in the first sum $i$ runs over even integers (odd terms will always be zero, as explained at the beginning of this section) and the second sum counts the ways to write  $i$ as the product of two integers $m_l$ and $n_l$, so $l$ runs from $1$ to the number of divisors of $i$, which we denote as $\text{div}(i)$. For large $i$, $\text{div}(i)$ is roughly of order $\sim \log i$. Similarly, for the coefficients $H^{(i)}_{m,n}$ of $q^i$ in $H_{m,n}$, we have
\be\label{eq:recursion2}
H^{(k)}_{m,n}=\sum_{i=2}^k\sum_{\substack{l=1\\m_ln_l=i}}^{\text{div}(i)}\frac{ R_{m_l,n_l}}{h_{m,n}+mn-h_{m_l,n_l}}H_{m_l,n_l}^{(k-i)}.
\ee
\begin{figure}
\centering
\includegraphics[width=0.4\textwidth]{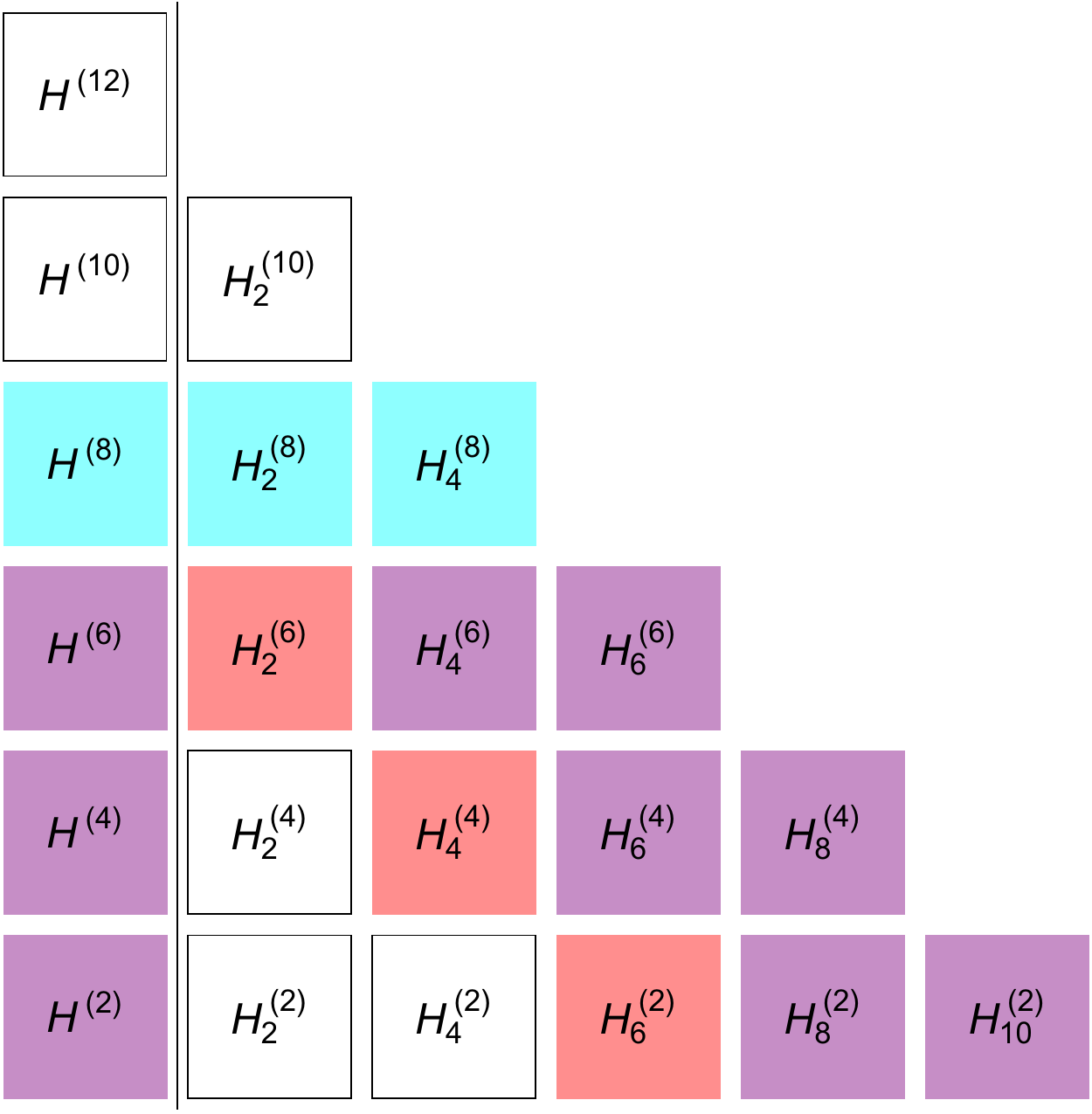}
	\caption{This figure shows a half-completed computation with max order $q^{12}$; each cell $H^{(k)}_i$ represents 2 to 4 distinct terms $H^{(k)}_{m_ln_l}$ with $m_ln_l=i$. The cyan row, order $q^8$, is currently being computed, and the red diagonal contains the terms which are being used in the computation of the cyan row. The purple cells have already been computed and are being stored for future use, and the white cells have not been computed yet or have been deleted to save RAM. The row with $k=0$ (which would be at the bottom) contains the seed terms $H^{(0)}=H^{(0)}_{m,n}=1$ and is not shown.}
\label{fig:HTriangle}
\end{figure}

Notice that in the above two equations,  $H^{(k)}$ and $H_{m,n}^{(k)}$ only depend on lower order terms $H_{m_l,n_l}^{(k-i)}$ for which $(k-i)+m_ln_l=k$. As illustrated in Figure \ref{fig:HTriangle}, we can perform the calculation from lower rows (small $k$) to upper rows (large $k$). 
In this way, when calculating $H_{m,n}^{(k)}$, all the $H_{m_l,n_l}^{(k-i)}$s are known already (and they are in the diagonal positions, which suggests to store them in diagonals), and there are only $\sim k \log k$ such terms, so the time complexity is only roughly $O\left(N^3 (\log N)^2\right)$. This is better than the literal implementation of the recursion relation (getting the coefficients $H^{(k)}$ by directly recursing down to $H_{m_l,n_l}^{(0)}$), which seems to have a complexity of $O(e^N)$.

There are several other tricks that one can do to even speed up the calculation. For example, one can precompute all of the residue 
prefactors $\frac{ R_{p,q}}{h_{m,n} + mn - h_{p,q}} \equiv C_{m,n,p,q}$ in \ref{eq:recursion2}. There are only 
$O(N\left(\log N\right)^2)$ of these, so we can save time by computing them in advance and reusing them. 
Although precomputation dramatically improves performance, it also doubles memory consumption; but since we store the $H_{m,n}^{(i)}$ in diagonals, this can be ameliorated by deleting them after 
they're used, as shown in \ref{fig:HTriangle}.

Precomputing $C_{m,n,p,q}$ can only improve overall speed if each of its terms can be computed in 
constant time. This is potentially problematic, since $R_{p,q}$ contains two products of $O(pq)$
complexity, but it can be solved by filling $R_{p,q}$ recursively -- $R_{p,q}$ can be computed in $O(p)$ time from $R_{p,q-2}$, and there are only $O(N \log N)$ of 
them, so the computational complexity of filling all $R_{p,q}$ is just $O(N^2 \log N)$. These can be further sped up by pairing up terms to rewrite all of 
the defining equations in terms of $b^2$ and $\lambda_{m,n}^2$ instead of $b$ and $\lambda_{m,n}$.
In addition to the reduced number of multiplications, this also allows the entire computation 
to be done using real numbers when $c > 25$, which is generally an order of magnitude faster. 
When $c < 25$, $b^2$ becomes complex, and even though the final coefficients must be real by 
unitarity, this only occurs at the very last step in the form of a 
$b^2 \leftrightarrow \frac{1}{b^2} = (b^2)^*$ symmetry.

We have implemented this algorithm in both Mathematica and C++(with Mathematica integration). The Mathematica notebook is included as a companion to this paper, while the C++ implementation  is maintained at \url{https://github.com/chussong/virasoro}. The C++ implementation is about one order of magnitude faster, and the coefficients used in this paper were obtained using it. The C++ implementation has used the GMP \cite{gnu:gmp}, MPFR \cite{gnu:mpfr}, MPC \cite{gnu:mpc}, and MPFR C++ \cite{gnu:mpfrcpp} numerical libraries. On standard personal computers we were able to compute the $H^{(k)}$ to $k=1000$ in around two minutes or $k=2000$ in about 22 minutes (for $c>25$ so that $b$ is real); the main barrier to going higher is memory consumption, which grows roughly as $N^3 \log N$: we need to remember $O(N^2 \log N)$ numbers and they need to be kept at $O(N)$ bits of precision due to the increasingly large cancellations between different $H_{m,n}$, which often reach into the thousands of binary orders of magnitude. 

Using a cluster with 128 GB of RAM, we estimate that we could reach order of $6000$ in a few hours. We also find that the coefficients of $q^i$ approach a power law in 
$i$ well before the limits of our desktop computation, and expect that a numerical fit for 
this power law would be good enough to get higher order coefficients.

At the end of this section, we want to mention an issue about the recursion relation if $b^2$ is a rational number. Notice that the denominator in \ref{eq:recursion2} and the denominator of $R_{m,n}$ in \ref{eq:Rmn} can be zero:
 \begin{align}
	h_{m,n} + mn - h_{m_l,n_l} &= 0\quad \Rightarrow \quad b^2=\frac{m+m_l}{n-n_l} \text{ or } \frac{m-m_l}{n+n_l}\\
	\lambda_{k,l} &=0\quad \Rightarrow \quad b^2=-\frac{k}{l}
	\end{align}
Both of these will eventually occur for any rational choice of $b^2$.  This would appear to preclude numerical computation entirely (since for numerical calculation, $b$ provided to the computer will always be rational), but actually for almost all rational numbers they will not appear until very high orders in the computation, so they can be ignored as long as the numerator or denominator of $b^2$ (as a irreducible fraction) is very large. In this paper, we've chose $\sqrt{c}$ to be irrational (and set $b$ to be a very high-precision number) to avoid this problem.

\section{Technical Details and Extra Plots}

\subsection{A Non-Perturbative Differential Equation for the Vacuum Block}
\label{app:DiffEqforVac}

Here we describe the functions appearing in the differential equation (\ref{eq:2ndOrderVacDiffEq}).  Note that although the equation itself is perturbative, its solution includes non-perturbative corrections to the heavy-light vacuum Virasoro block.
The equation was derived \cite{Fitzpatrick:2016ive} by studying the general  differential equations satisfied by degenerate operators and then analytically continuing these equations in the integer index $r$ labeling the degenerate operators.  We should also note that although equation (\ref{eq:2ndOrderVacDiffEq}) only includes some of the first $1/c$ corrections, if one zooms in on the vicinity of the forbidden singularities by holding $\sqrt{c} (z-z_n)$ fixed at large $c$, then the equation incorporates all of the leading effects at large $c$.  As discussed in \cite{Fitzpatrick:2016ive}, there are both general arguments and consistency checks on the validity of equation (\ref{eq:2ndOrderVacDiffEq}).  

\begin{figure}
\centering{}\includegraphics[width=0.6\textwidth]{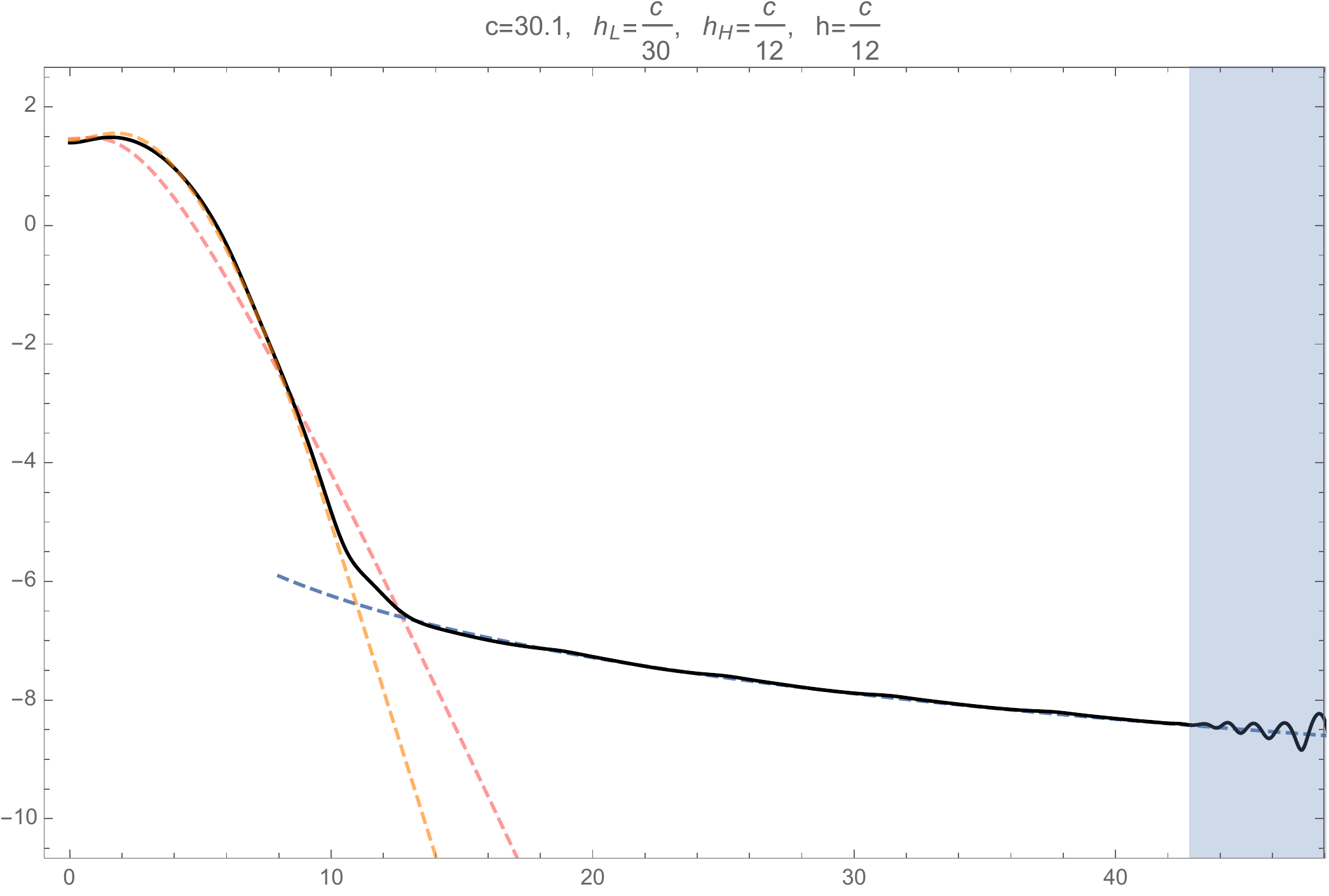}
\caption{This figure corresponds to the top-right plot of figure \ref{NonVBlock}, but includes a match to the semiclassical blocks obtained using the methods of \cite{Fitzpatrick:2016mjq}, which allow for $h, h_L \propto c$.  The poorly fitted dashed line is the approximation of equation (\ref{eq:HeavyLightBlocks}), which assumes $h, h_L \ll c$, and clearly provides a much less reliable fit for these parameter values.}
\label{fig:NonVacSemiVSFull}
\end{figure}

\begin{figure}[h]
\begin{centering}
\includegraphics[width=0.99\textwidth]{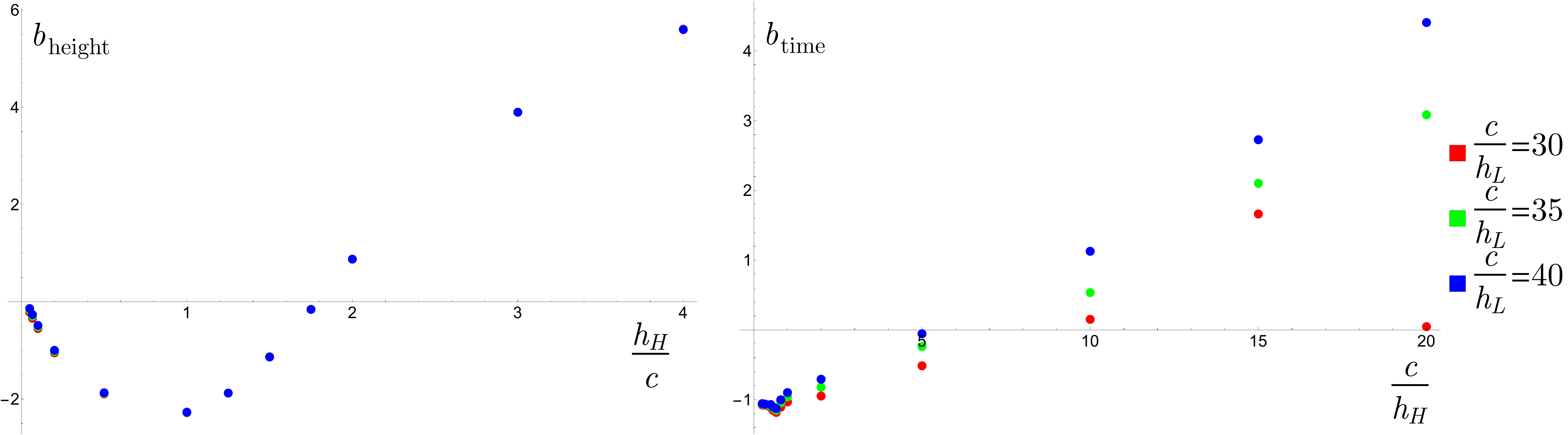}
\caption{ We have found empirically that the time and height of the maxima of heavy-light Virasoro blocks have a simple dependence on both $h$ and $h_H$.  This figure shows data on the parameters $b_\text{height}$ and $b_\text{time}$ defined in equations (\ref{eq:Vmaxdef}) and (\ref{eq:tmaxdef}).  These plots both have $c=10$. Each point is obtained from linear fitting of data points at $\frac{h}{c} = \frac{n}{3}$ for $n = 1, 2, \cdots, 30$.    We see explicitly that there is very little dependence on $h_L$, especially at large values of $h_H$. }
\label{fig:bparameterdata}
\end{centering}
\end{figure}

We identify the parameter $r = 2 \pi i T_H = \sqrt{1 - \frac{24 h_H}{c}}$, so that $T_H$ is the Hawking temperature associated with the heavy operator.  We also are using a Euclidean time variable $\tau = - \log(1-z)$. Then the functions included in equation (\ref{eq:2ndOrderVacDiffEq}) are $g_H \equiv g_{2 \pi i T_H}$ with
\be
g_r(\tau) &=& \coth \left( \frac{\tau}{2} \right) - r \coth \left( \frac{r \tau}{2} \right)
\ee
and $\Sigma_H \equiv \Sigma_r + \Sigma_{-r}$ where we define
\be
\Sigma_r(\tau) &=&
-\frac{1}{r \sinh \left( \frac{r \tau}{2} \right)} \left( e^{- \frac{ r \tau}{2}} \tilde{B}_r(\tau) + e^{\frac{r \tau}{2}} \tilde{B}_r(-\tau) -2 \cosh \left( \frac{r \tau}{2} \right) \tilde{B}_r(0) \right).
\ee
Finally, we have introduced the function $\tilde{B}_r(t)$ which can be represented as
\be
\tilde{B}_r(\tau)  &=& -\log(1-e^{\tau}) - \frac{e^{ r \tau} {}_2F_1(1,r,1+r,e^\tau)}{r} 
 \label{eq:UpsHyp}
\ee
For derivations and more complete descriptions see \cite{Fitzpatrick:2016ive}.

\subsection{Some Extra Plots}

\begin{figure}[h]
\begin{centering}
\includegraphics[width=0.49\textwidth]{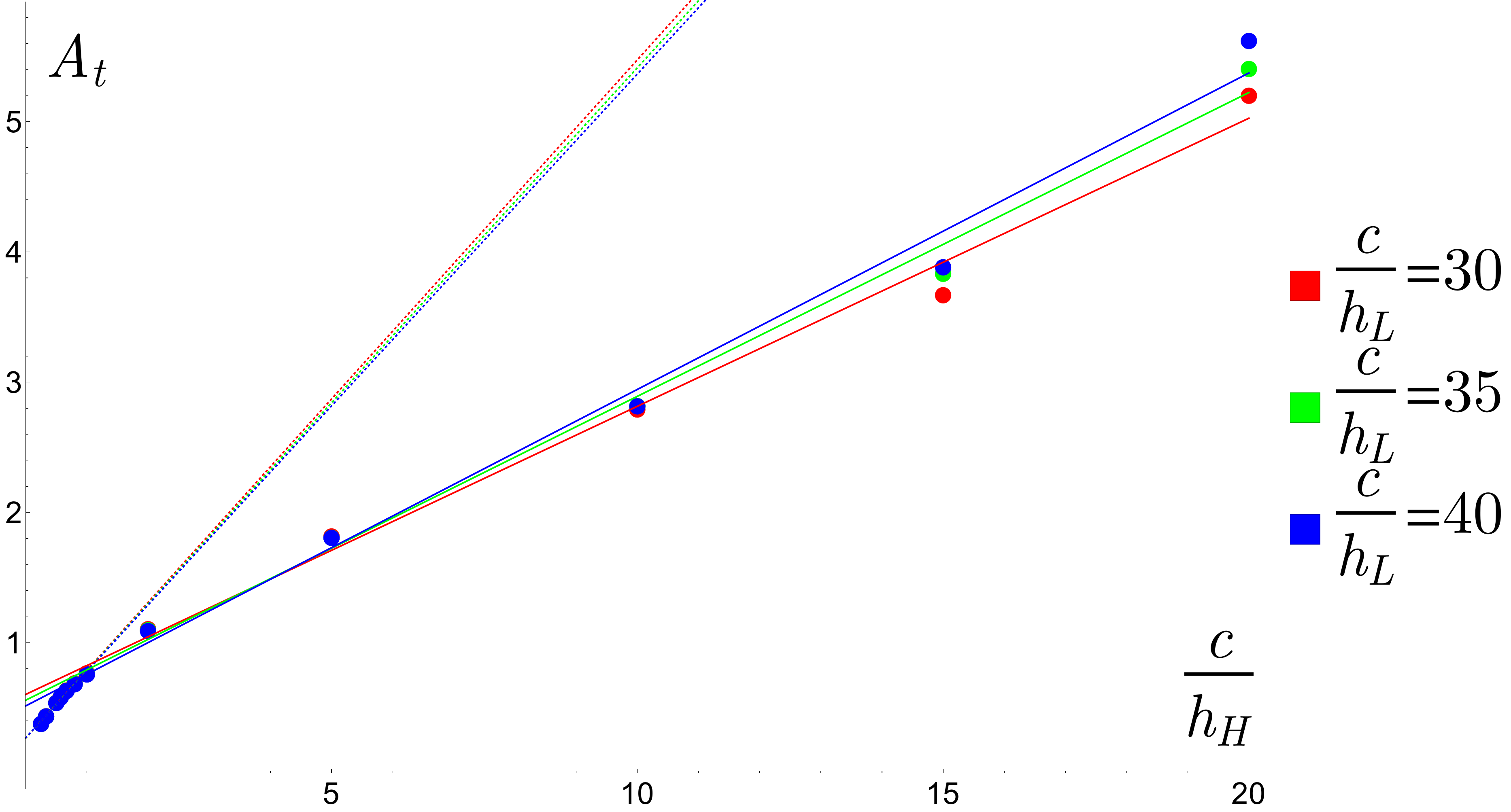}
	\caption{ This is a version of figure \ref{fig:aparameterfits} where we have zoomed out to show the small $\frac{h_H}{c}$ region. The zoomed-out points with $\frac{c}{h_L} = (30, 35, 40)$ more closely fit slopes $(0.221, 0.233, 0.242)$, which are shown as solid lines; the $(0.521, 0.515, 0.509)$ fits for large $\frac{h_H}{c}$ are shown as dotted lines.  }
\label{fig:at_large_hh}
\end{centering}
\end{figure}

In this section we have included some extra plots for readers who might like to some more details and examples.  These include  the semiclassical fit to our numerical results for $h, h_L \propto c$ using \cite{Fitzpatrick:2016mjq} (figure \ref{fig:NonVacSemiVSFull}), the behavior of the $b_\text{time}$ and $b_\text{height}$ parameters from equations (\ref{eq:Vmaxdef}) and (\ref{eq:tmaxdef}) (figure \ref{fig:bparameterdata}) and a version of figure \ref{fig:aparameterfits} zoomed in on the large $h_H/c$ region (figure \ref{fig:at_large_hh}), which is rather compressed in that figure.

\begin{figure}[h]
\begin{centering}
\includegraphics[width=0.99\textwidth]{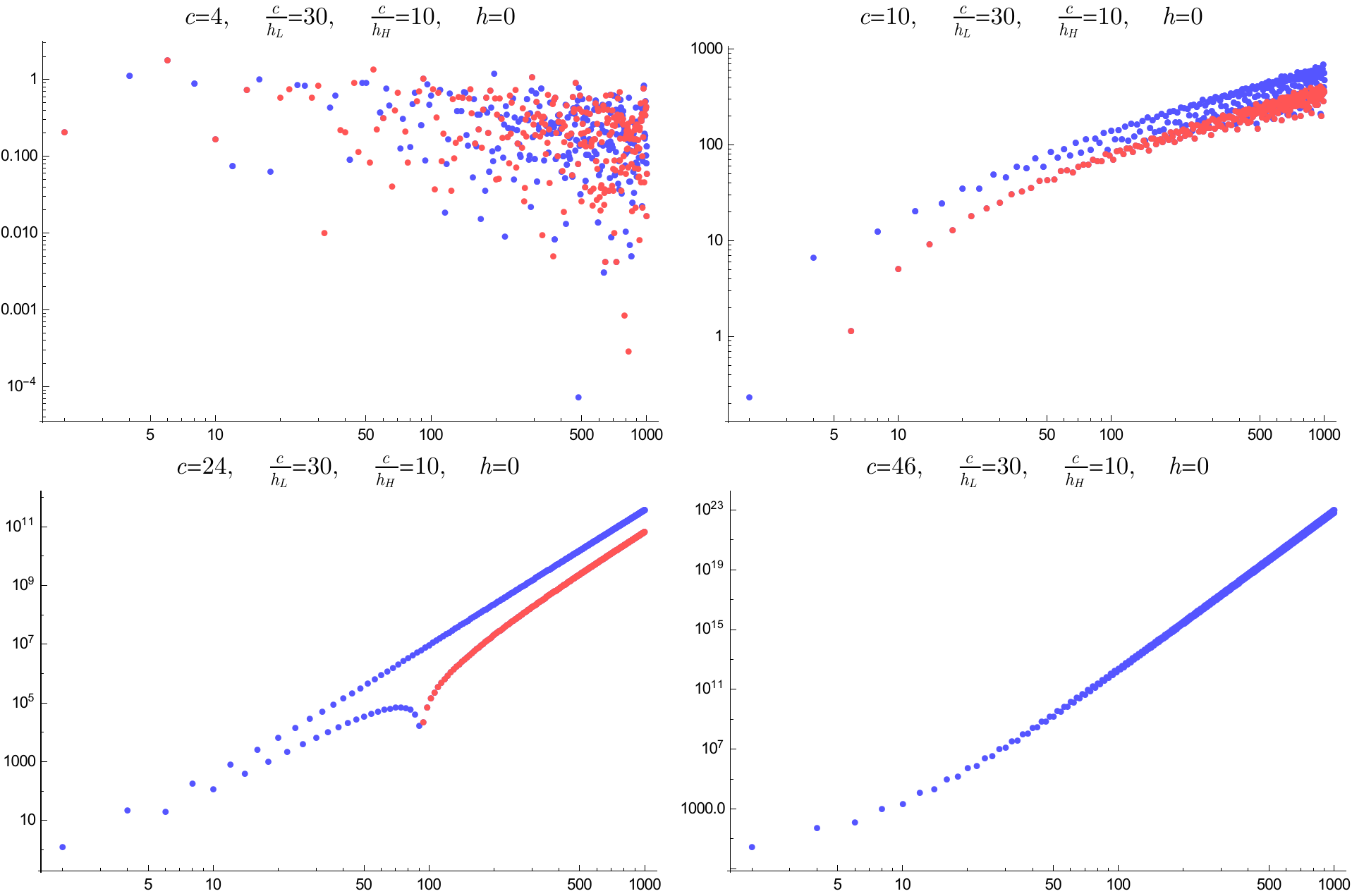}
\caption{ This figure shows the coefficients $c_{n}$ of the $q^{2n}$ expansion of $H$.  We plot $|c_{n}|$ as a function of $n$, with both $n$ and $c_{n}$  on $\log$ scales, for increasing $c$ with $\frac{h_L}{c}$ and $\frac{h_H}{c}$ held constant. The sign of the $c_{n}$ are illustrated by the color of the points, with blue for positive coefficients and red for negative coefficients.}
\label{fig:CoeffEvolution}
\end{centering}
\end{figure}

We also show some plots of the more complicated coefficient behavior which was 
alluded to in section \ref{sec:CoeffPowerLaw}, with the sign of the coefficients corresponding to the color of plotted points. Figure \ref{fig:CoeffEvolution} illustrates a very common scenario where the coefficients are chaotic at low $c$, but as $c$ increases they coalesce into distinct positive and negative lines. A  spike-shaped feature then appears at low order and moves upward, turning the coefficients that it passes positive, until all (visible) coefficients have become positive. The two lines then gradually merge into a single power law similar to those shown in figure \ref{fig:Coeffpower}.

\bibliographystyle{utphys}
\bibliography{VirasoroBib}

\end{document}